\documentclass[unsortedaddress, reprint, amsmath,amssymb,amsfonts,prb,aps, showpacs]{revtex4-1}
\usepackage{graphicx}
\usepackage{amsmath}
\usepackage{amssymb}
\usepackage{subfigure}
\usepackage{color}
\usepackage{array}
\usepackage{bm}
\begin{document}
\title{Decoherence wave in magnetic systems and creation of N\'eel antiferromagnetic state by measurement}

\author{Hylke C. Donker}
\email{h.donker@science.ru.nl}
\affiliation{Radboud University, Institute for Molecules and Materials, Heyendaalseweg 135, NL-6525AJ Nijmegen, The Netherlands}
\author{Hans De Raedt}
\affiliation{Department of Applied Physics, Zernike Institute for Advanced  Materials, University of Groningen, Nijenborgh 4, NL-9747AG Groningen, The Netherlands}
\author{Mikhail I. Katsnelson}
\affiliation{Radboud University, Institute for Molecules and Materials, Heyendaalseweg 135, NL-6525AJ Nijmegen, The Netherlands}


\begin{abstract}
The interplay between the singlet ground state of the antiferromagnetic Heisenberg model and the experimentally measured N\'eel
state of antiferromagnets is studied. To verify the hypothesis [M. I. Katsnelson \emph{et al.}, Phys. Rev. B 63, 212404 (2001)]
that the latter can be considered to be a result of local measurements destroying the entanglement of the quantum ground state,
we have performed systematic simulations of the effects of von Neumannmeasurements for the case of a one-dimensional
antiferromagnetic spin-1/2 system for various types and degrees of magnetic anisotropies. It is found that in the ground state,
a magnetization measurement can create decoherence waves [M. I. Katsnelson \emph{et al.} Phys. Rev. A 62, 022118 (2000)] in the
magnetic sublattices, and that a symmetry breaking anisotropy does not lead to alignment of the spins in a particular direction.
However, for an easy-axis anisotropy of the same order magnitude as the exchange constant, a measurement on the singlet ground state can create N\'eel-ordering in finite systems of experimentally accessible size.
\end{abstract}

\maketitle
\section{Introduction}
Magnetism has played a crucial role in the development of electronic data storage. To keep up with the ever increasing demand of storage space, finding new technologies which are able to do so is paramount. This requires fundamental insight in the workings of magnetism at the smallest possible length scales. Recent development of scanning probe microscopy allows one to do precisely this, namely study magnetic particles atom-by-atom \cite{KHAJ11,KHAJ13}. Apart from technical perspectives, this also opens a new way to study fundamental issues of the quantum physics of magnetism.

At present, the origin of magnetically ordered states is well understood from basic principles of quantum physics~\cite{VONS74,BLUN01,WHIT07}.
Nevertheless, some subtle points of fundamental importance still seem to require deeper understanding. The origin of the antiferromagnetic N\'eel state is one of them \cite{VONS74,IRKH86}. Neutron diffraction experiments seem to suggest the existence of sublattice magnetization in antiferromagnetic materials~\cite{KITT04}, even for one dimensional systems~\cite{TENN95, KOJI97}. Therefore the conventional picture of an antiferromagnetic material in the low temperature ordered phase is a N\'eel state in which neighbouring spins are anti-parallel, i.e. $|\psi_N\rangle = |\uparrow \downarrow \uparrow \dots\rangle$ or $|\psi_{N^\prime}\rangle = | \downarrow \uparrow \downarrow \dots\rangle$. A more detailed analysis~\cite{IRKH86} shows that these basic observations, as well as most other experimental manifestations of antiferromagnetism, can formally be described without broken symmetry and sublattices. What is required is long-range order of N\'eel type in the sense that there are singularities in the spin pair correlation functions~\cite{IRKH86}.

To stress, the aforementioned N\'eel state is not the ground state of the antiferromagnetic Heisenberg Hamiltonian (HH): there is only a partial overlap with the ground state (see e.g. Ref.~\onlinecite{BROC14}). Only in the limiting case in which the product of the spin, $S$, and atom co-ordination number, $z$, tends to infinity ($1/(zS) \rightarrow 0$) does the energy of the N\'eel state coincide with the ground state energy of the HH~\cite{VONS74,IRKH86}. 

One manifestation of the difference between $|\psi_N\rangle$ and the ground state $|\psi_0\rangle$ is in the sublattice magnetization. The magnetization operator $\bm{S}_A$ of sublattice A (or equivalently $\bm{S}_B$ of sublattice B) does not commute with the Hamiltonian. Therefore the sublattice magnetization is not a good quantum number~\cite{VONS74}. In fact, 
as is well-known, in one dimension the ground state of the antiferromagnetic HH with nearest-neighbour interactions and periodic boundary conditions is a non-degenerate singlet (i.e. $S$=0)~\cite{MARS55, LIEB61, LIEB62, YANG66}. Hence, in the ground state the sublattice magnetization vanishes~\cite{PRAT61}.

In order to bridge the gap between the experimentally measurable sublattice magnetization and the ground state singlet configuration of the antiferromagnetic HH, one can introduce a conjugate field.
The textbook procedure is to introduce an infinitesimal staggered magnetization $\bm{h}_{st}$, which breaks time reversal symmetry~\cite{HUAN00, BLUN01, MAJL07, KUZE10}. The conjugate field $\bm{h}_{st}$ points in a particular direction, e.g. the $z$-direction, and alternates in sign when going from one sublattice to an other. In contrast to ferromagnetic systems, for antiferromagnetic systems there is no clear physical picture to which this staggered magnetization should correspond to~\cite{IRKH86, KUZE10}.

In one dimension the discrepancy between the N\'eel state and the ground state of the HH is especially large due to the small co-ordination number. One dimensional antiferromagnetic materials (such as the (isotropic) Heisenberg chains KCuF$_3$~\cite{NAGL91, TENN95} and Sr$_2$CuO$_3$~\cite{MOTO96, KOJI97}) have already been known for some time. However, these systems are not well suited to measure the magnetization at individual sites.

New experimental techniques allow the creation of artificial spin chains in which spins can be individually probed: this can be done using spin polarized STM techniques~\cite{HIRJ06, LOTH12, KHAJ12, HOLZ13, SPIN14}, chains of trapped ions~\cite{FRIE08,RICH14} or optical lattices~\cite{TROT08, SIMO11}. This has caused renewed interest in the ground state of antiferromagnetic one dimensional spin systems. In particular it was claimed that singlet~\cite{HIRJ06} and N\'eel configurations~\cite{LOTH12, KHAJ12} can in fact be measured. Therefore it is of interest to see how the N\'eel state can emerge from the ground state. It was suggested that the formation of sublattice magnetization can be induced by the act of local measurements~\cite{KATS01}. However, the supporting analytical calculations in that work were based on the trial wave function~\cite{IRKH86} which is accurate only in the $1/(zS) \rightarrow 0$ limit, and therefore it is expected that this is a poor approximation for a one dimensional spin-1/2 system.

In the present paper the emergence of the N\'eel state from the ground state $|\psi_0\rangle$ is studied by analyzing the effect of a measurement by means of straightforward 
(numerically exact) computation of the time-dependent Schr\"odinger equation.
We consider a ring of spin-1/2 particles, see Fig.~1. 
The effect of a localized (i.e. a single site) measurement on the system in the ground state is analyzed
and the influence of the anisotropy on the result of the measurement is studied.

\section{Formulation of the model}
\begin{figure}
\includegraphics[scale=0.2]{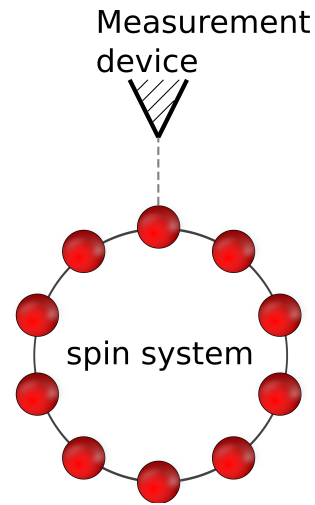}
\caption{(Color online) Schematic of the measurement set-up. The system consists of $N$ spin-1/2 particles with nearest-neighbour interactions. The spins are arranged in a ring and the system is prepared in the ground state. Measurement is performed at a single site.}
\label{fig:setup}
\end{figure}
According to von Neumann~\cite{NEUM55} a measurement can be described as the (non-unitary) transition from a pure to a mixed state $\rho \rightarrow \rho^\prime = \sum_i P_i \rho P_i$, where $\rho$ is the density matrix and $P_i$ are (idempotent) projection matrices summing to unity. This process can be described in various ways~\cite{ZURE03, JOOS13, NIEU13}, but for simplicity we restrict ourselves to idealized local instantaneous magnetization measurements, that is, single particle measurements in which the transition from pure to mixed state is immediate. It has been shown that in a closed system a local measurement can induce a propagating disturbance, a so-called decoherence wave~\cite{KATS00, KATS01, HAMI05}. Up to now, the calculations were done only for simple exactly solvable systems such as the ideal (or weakly non-ideal) Bose gas~\cite{KATS00} or the one-dimensional Ising model in a transverse field~\cite{HAMI05}.

In the case of spin-1/2 systems, as considered here, an instantaneous magnetization measurement along Cartesian axis $\alpha$ on spin $m$ corresponds to application of the projection operator
\begin{equation}\label{eq:projector}
P^{\pm \alpha}_m = \frac{1 \pm 2{S}^\alpha_m}{2} \, ,
\end{equation}
to the wave function. Here and in the following, ${S}^\alpha_m$ is the spin operator for site $m$ along Cartesian axis $\alpha$. The + (-) sign of the projection operator indicates projection parallel (anti-parallel) to axis $\alpha$. Throughout this article units in which $\hbar=1$ are used.

Subject of the present study is the effect of a von Neumann measurement on the ground state of the one-dimensional spin-1/2 Heisenberg Hamiltonian (HH), see Fig.~\ref{fig:setup}.
The Hamiltonian is given by~\cite{VONS74,WHIT07}
\begin{equation}\label{eq:HH}
H_0 = J \sum_{<i,j>} \bm{S}_i \cdot \bm{S}_j \, ,
\end{equation}
where $\langle i, j \rangle$ denotes pairs of nearest neighbours and $J$ is the exchange parameter. Henceforth AFM ($J> 0$) finite systems of an even number of $N$ spins are considered, with periodic boundary conditions, i.e. $\bm{S}_{i+N} = \bm{S}_{i}$.

The effect of symmetry on the formation of the N\'eel state can be examined by introducing an anisotropy $H^{\prime}$ of strength $\Delta$. Specifically, anisotropies of the form
\begin{equation}\label{eq:H_aniso}
H^{\prime} =  \Delta \sum_{<i,j>} S^z_i S^z_{j} \, ,
\end{equation}
will be studied such that the total Hamiltonian takes the form $H=H_0 + H^\prime $.
Note that the anisotropic interaction $H^{\prime}$ preserves time-reversal symmetry and that the ground state of the Hamiltonian $H$ is non-degenerate for arbitrary $\Delta$~\cite{YANG66}.

Observables such as the magnetization of site $l$ in direction $\beta$ after a projection of Eq.~(\ref{eq:projector}) can be calculated using:
\begin{equation}\label{eq:magn}
\left\langle S^\beta_l(t) \right\rangle = \mathrm{Tr} \left[ S^\beta_l(t)  \frac{P^{\pm \alpha}_m \rho_0 P^{\pm \alpha}_m}{N_0}  \right] \, ,
\end{equation}
where $\rho_0$ is the density matrix~\cite{FANO57} of the ground state and $N_0$ is a normalization factor to insure that $\rho = P^{\pm \alpha}_m \rho_0 P^{\pm \alpha}_m/N_0$ has unit trace. Similar relations can be constructed for e.g. the equal time correlation function.

\section{Simulation procedure}
\begin{figure}
\includegraphics[scale=0.45]{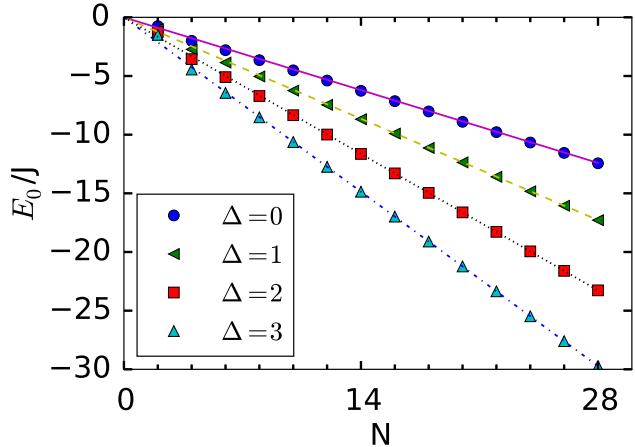}
\caption{(Color online) The ground state energy $E_0$ as a function of the chain length N. The markers indicate the values calculated using the Lanczos algorithm~\cite{GOLU96} and the slope of the (straight) lines follow from the (exact) Bethe Ansatz solution in the $N\rightarrow \infty$ limit~\cite{ORBA58}.}
\label{fig:E_gs}
\end{figure}

To compute the ground state of the Hamiltonian $H$ we use the Lanczos algorithm~\cite{GOLU96}.
The unitary time evolution of the wave function $|\Psi \rangle$, or equivalently the  evolution of the density matrix $\rho = |\Psi \rangle \langle \Psi|$,  is calculated using the Chebyshev polynomial expansion, which yields numerically exact results up to machine precision~\cite{DOBR03,RAED06}.

As a consistency check, in Fig.~\ref{fig:E_gs} the calculated values of the ground state energy are compared with the exact result from the Bethe Ansatz in the thermodynamic limit~\cite{ORBA58}. Figure~\ref{fig:E_gs} shows that there is excellent agreement between ground state energy of the finite $N$ calculation and the Bethe Ansatz in the thermodynamic limit.
In addition, both the calculated ground state and time evolution has been cross-checked with exact diagonalization for small values of $N\leq$ 12. In all simulations, the ground state shows zero magnetization (as required for a singlet) which corroborates the correctness of the calculated ground state.

Most simulation results presented here are for three different chain lengths namely, $N$=10, 20 and 28. This choice is motivated by the small size of systems in trapped ions~\cite{RICH14,MURM15} and spin polarized STM~\cite{LOTH12,KHAJ12} experiments on the one hand, and the role of finite size effects and the computational complexity (Hilbert space grows as 2$^N$) on the other hand.

\section{Results}
\begin{figure*}
\centering
\subfigure[N=10, m even.]{\label{fig:N10_isotropic_single_even}
\includegraphics[width=0.3\textwidth]{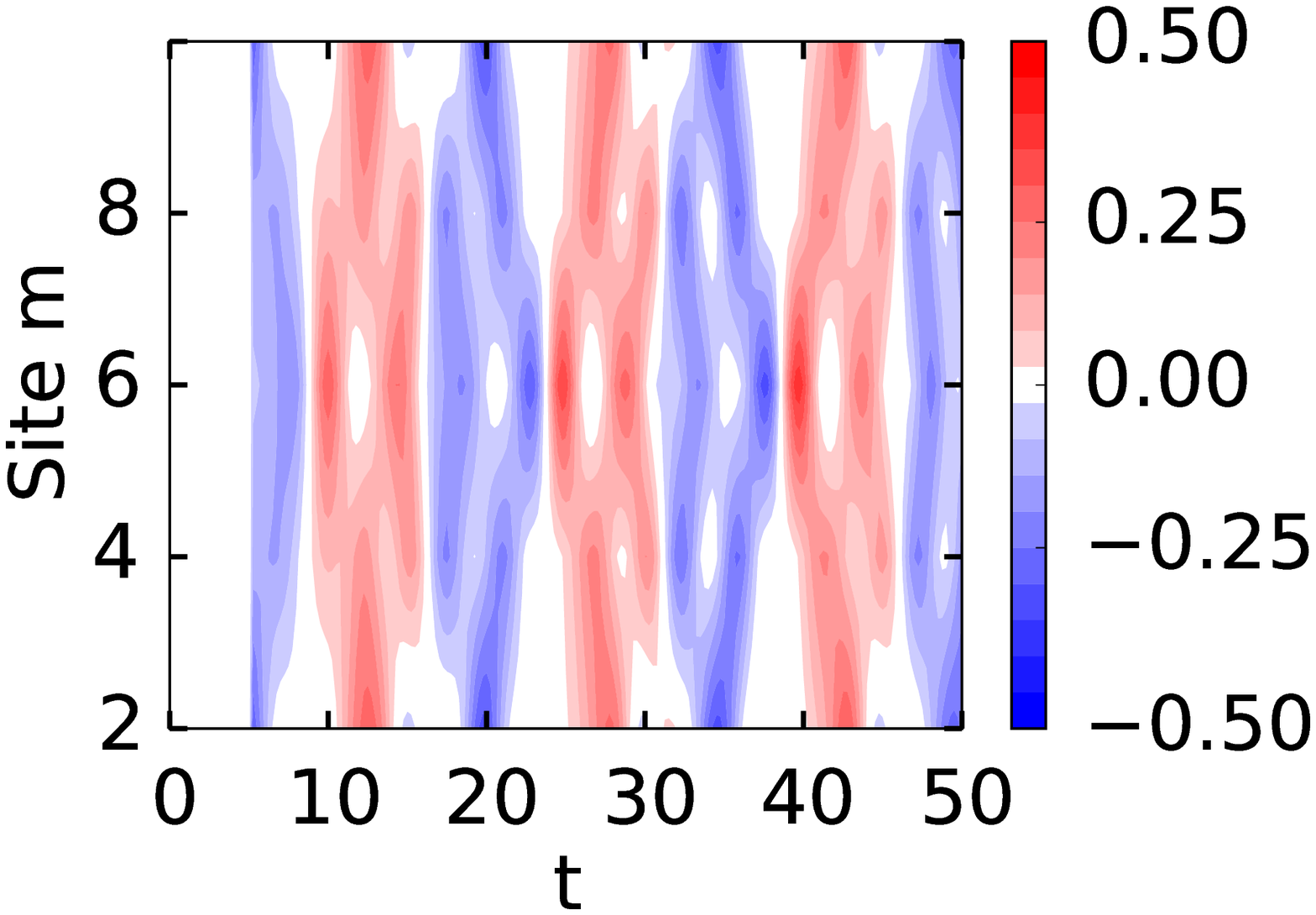} }
\subfigure[N=20, m even.]{\label{fig:N20_isotropic_single_even}
\includegraphics[width=0.3\textwidth]{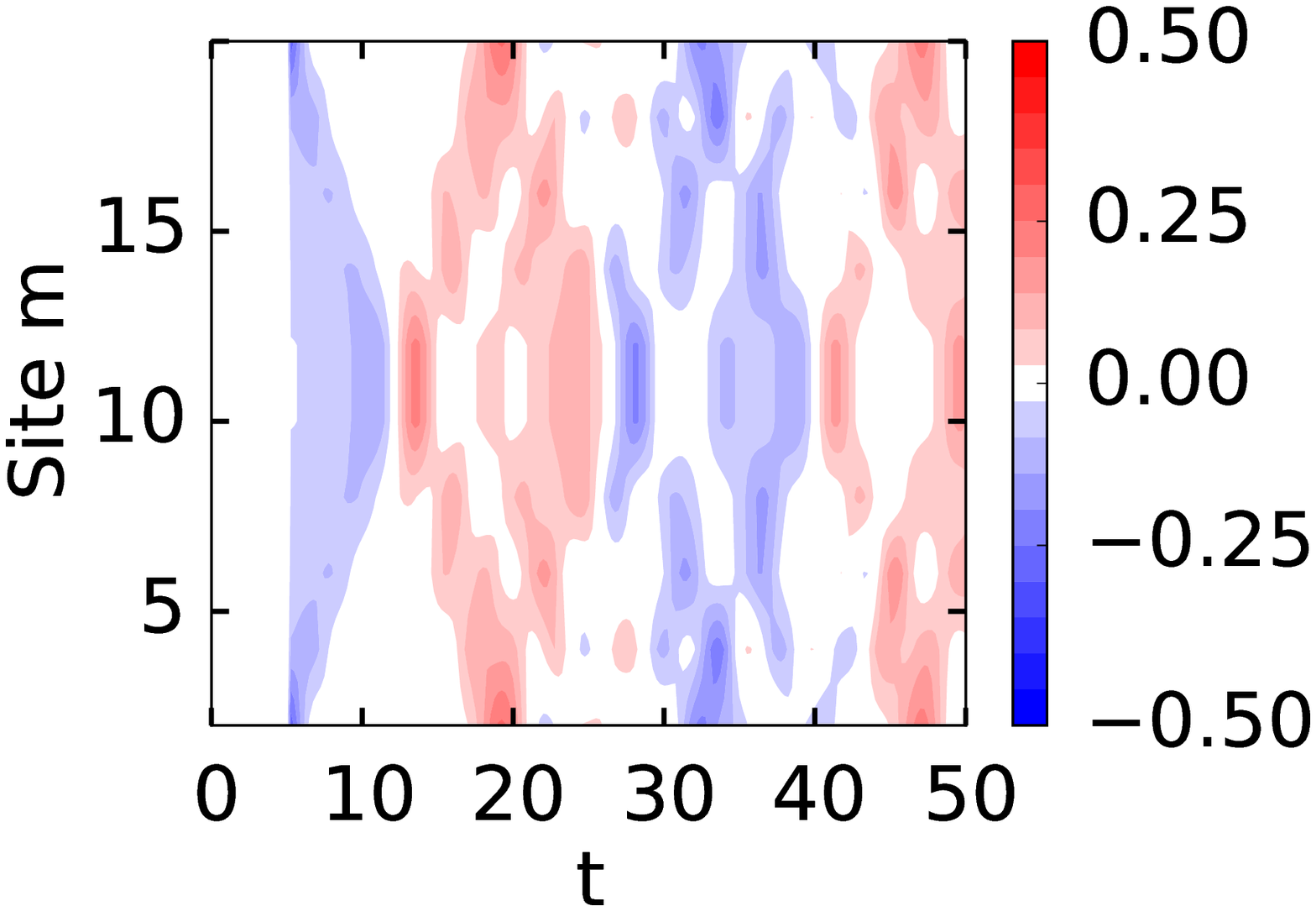} }
\subfigure[N=28, m even.]{\label{fig:N28_isotropic_single_even}
\includegraphics[width=0.3\textwidth]{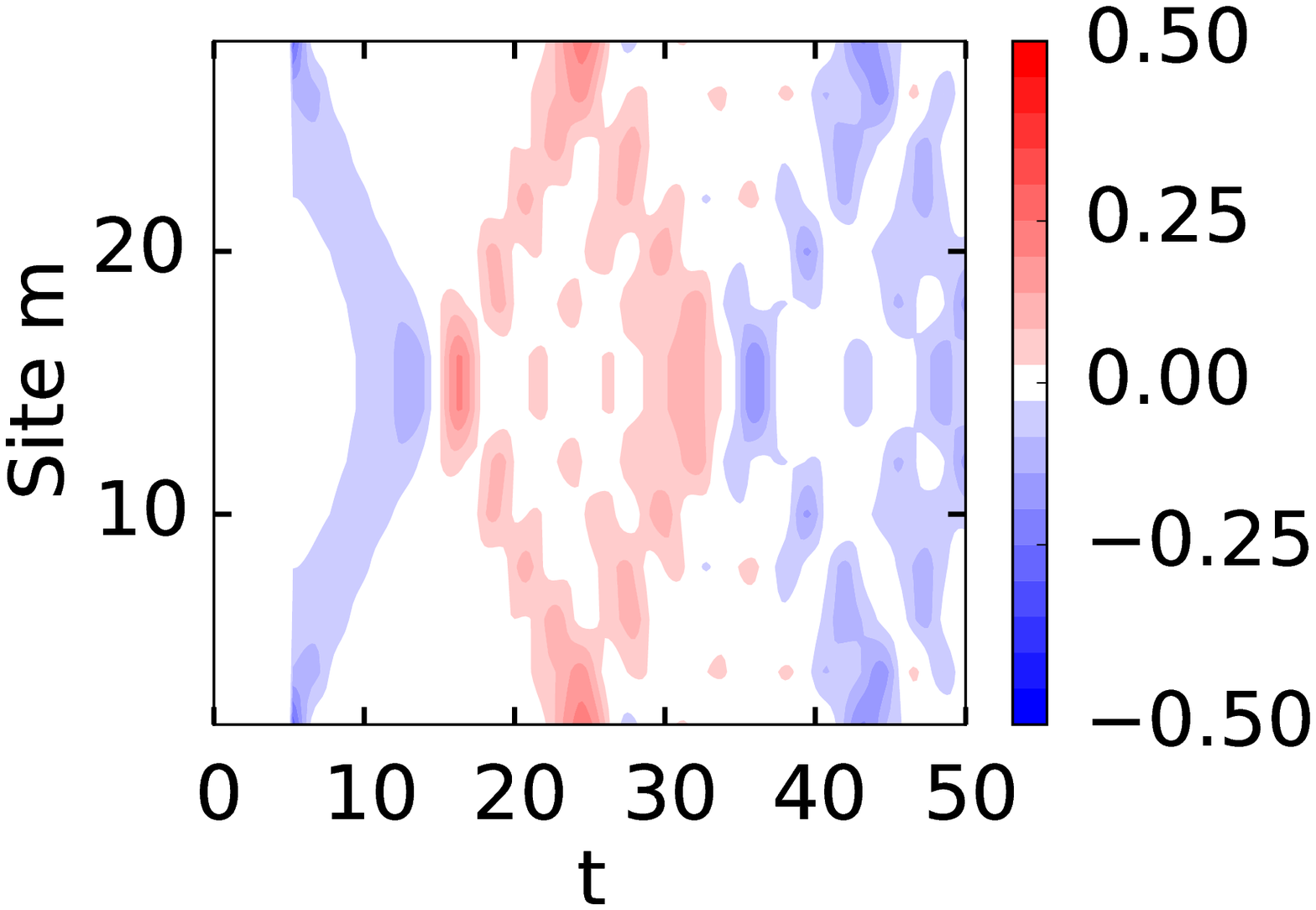} }
\subfigure[N=10, m odd.]{\label{fig:N10_isotropic_single_odd}
\includegraphics[width=0.3\textwidth]{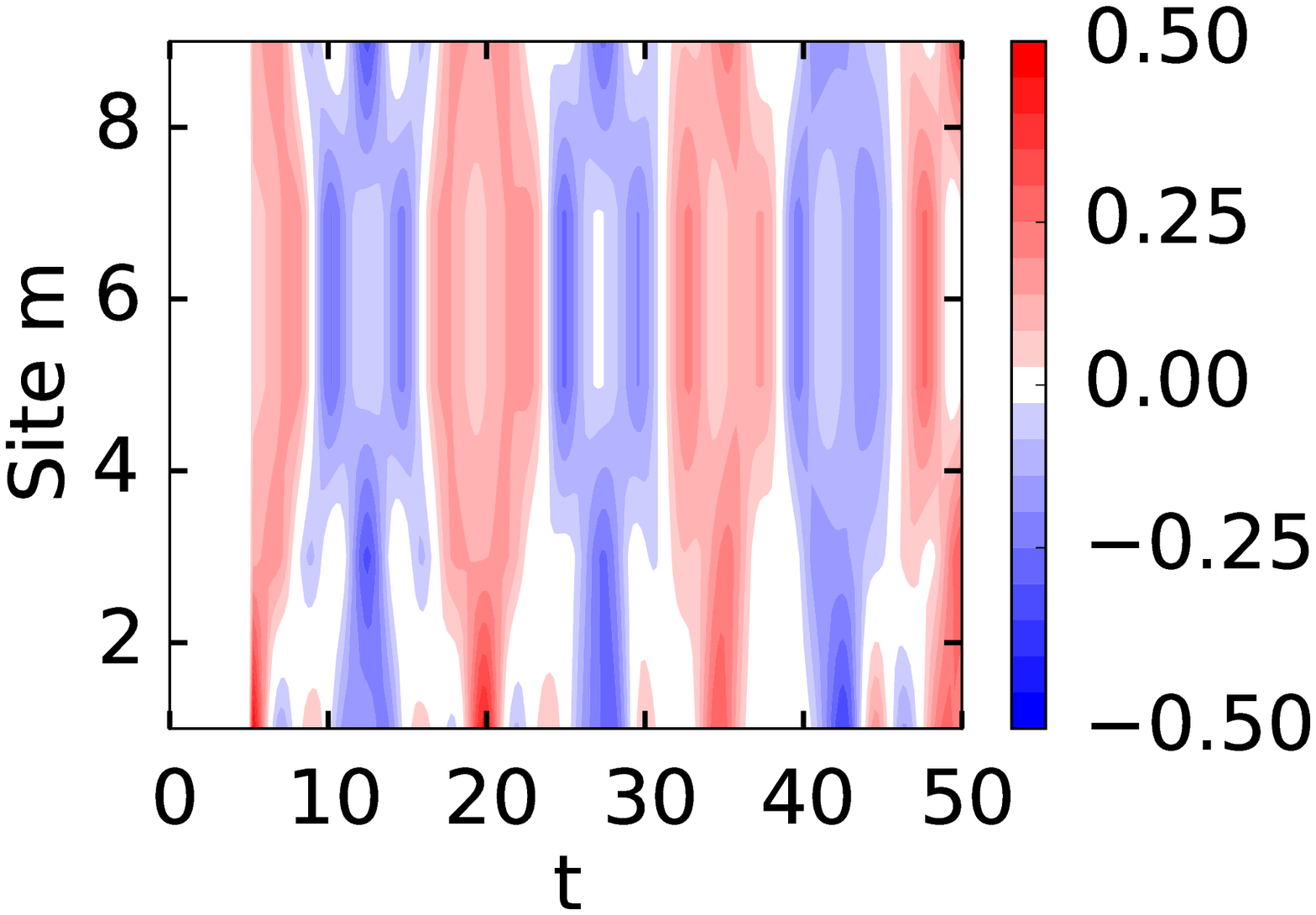} }
\subfigure[N=20, m odd.]{\label{fig:N20_isotropic_single_odd}
\includegraphics[width=0.3\textwidth]{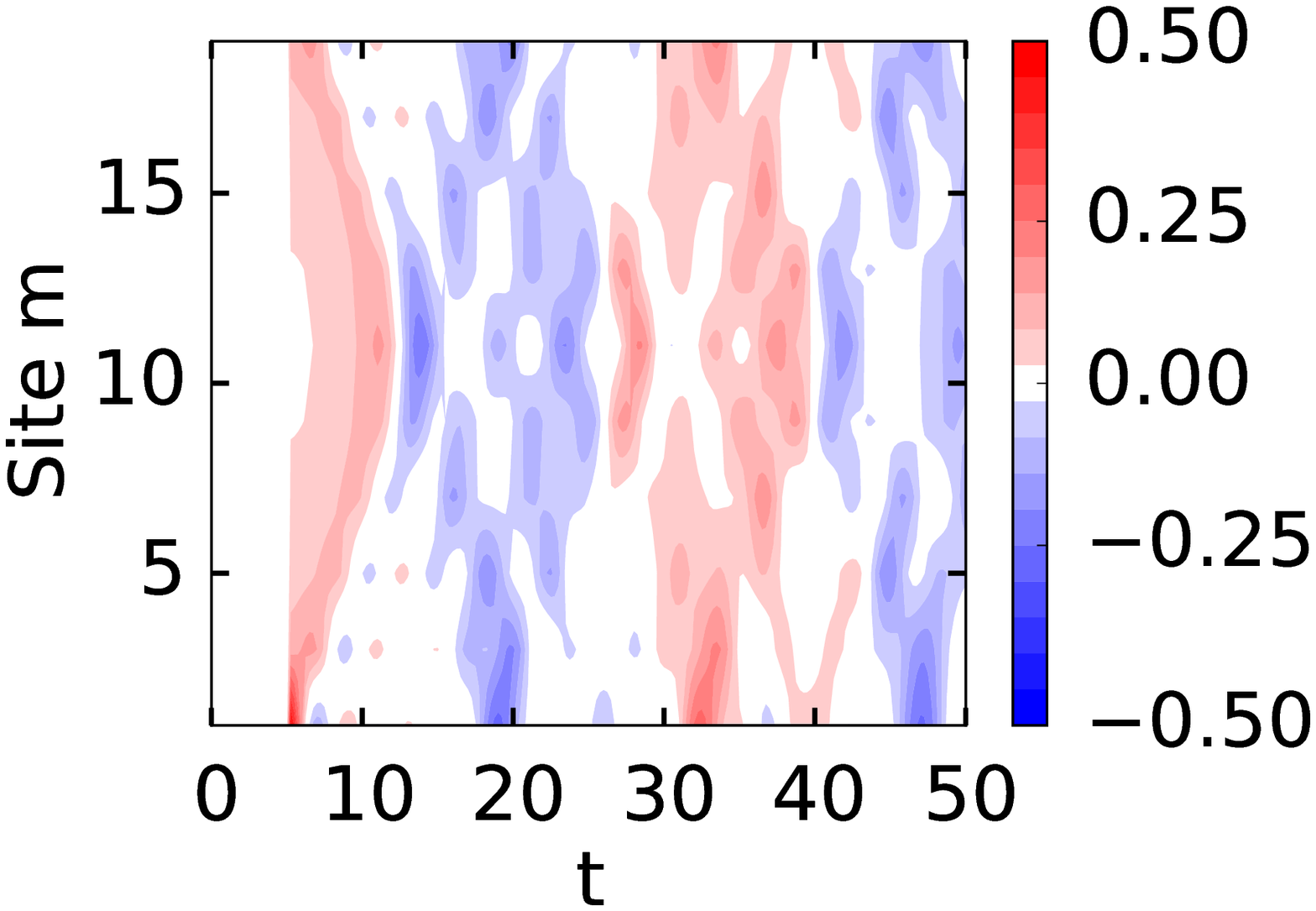} }
\subfigure[N=28, m odd.]{\label{fig:N28_isotropic_single_odd}
\includegraphics[width=0.3\textwidth]{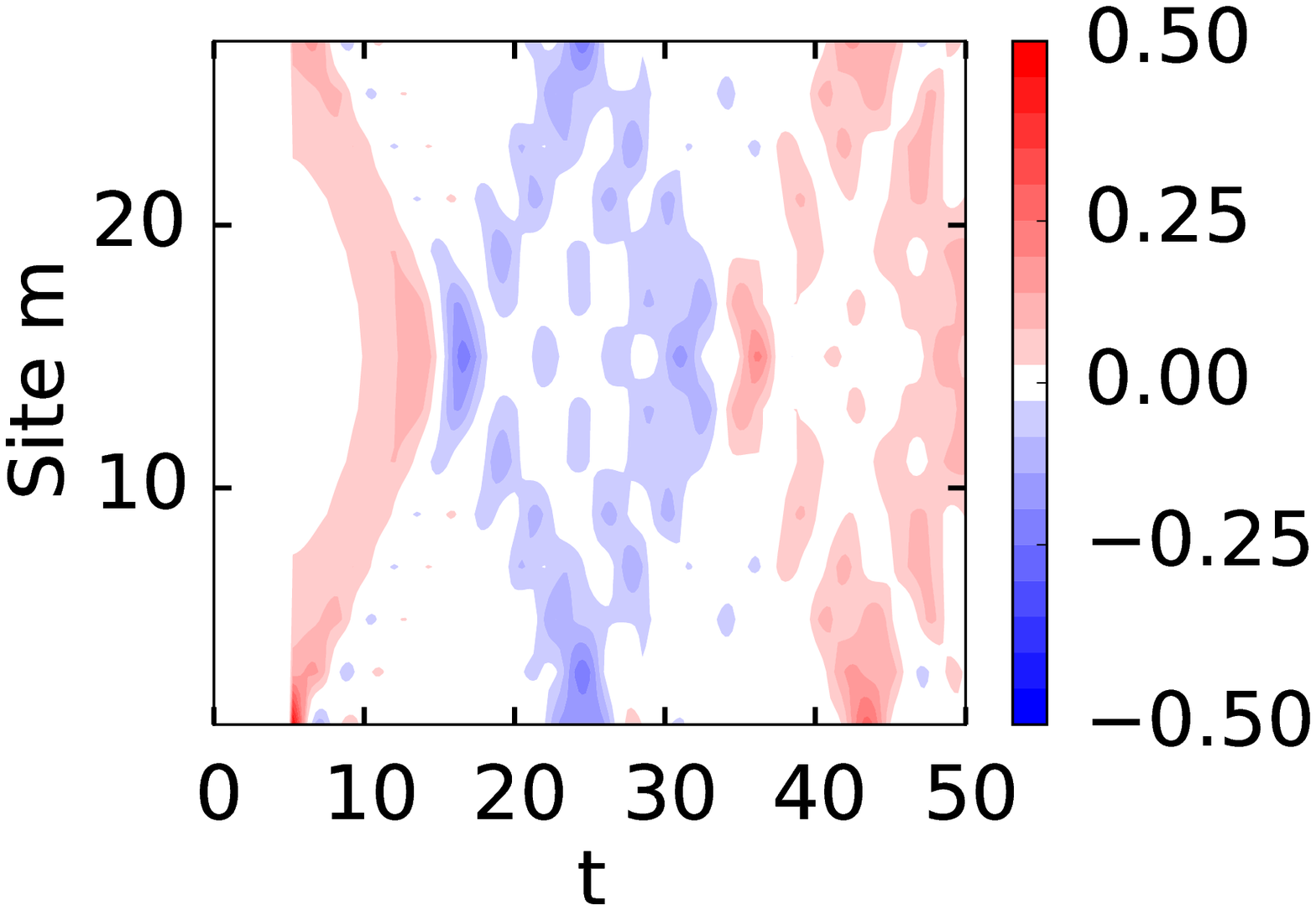} }
\caption{(Color online) Time evolution of the magnetization $\langle S_m^z (t) \rangle$
for the isotropic (i.e. XXX) AFM Heisenberg spin chain of length $N$.
The system at $t = 0$ is prepared in the ground state after which at $t = 5$ spin 1 is projected on the +$z$-axis.
}
\label{fig:heisenberg_isotropic_single}
\end{figure*}

\subsection{Single measurement}
The isotropic (or XXX) HH, i.e. $\Delta = 0$, will be considered first.
In Fig.~\ref{fig:heisenberg_isotropic_single} the single site magnetization $\langle S^z_{m} \rangle$ ($m = 1 \dots N$)
is plotted as function of time $t$ for different values of the chain length $N$.
Time $t$ has been made dimensionless, i.e. $t \rightarrow tJ\hbar$, throughout this article.
For clarity of presentation the magnetization is split up in even (panels \ref{fig:N10_isotropic_single_even}, \ref{fig:N20_isotropic_single_even}, and \ref{fig:N28_isotropic_single_even}) and odd sites (panels \ref{fig:N10_isotropic_single_odd}, \ref{fig:N20_isotropic_single_odd}, and \ref{fig:N28_isotropic_single_odd}).
In the classical N\'eel picture the separation of even and odd sites would correspond to magnetic sublattices. At time step $t = 5.0$ spin 1 is projected on to the positive $z$-direction (Eq.~(\ref{eq:projector})).  Figures depicting the magnetization in the $x$- and $y$-direction have been omitted since no significant deviations from zero could be observed.

Fig.~\ref{fig:heisenberg_isotropic_single} illustrates that a measurement induces four decoherence waves~\cite{KATS00, KATS01, HAMI05}: in each sublattice a forward and backward evolving wave is created. For $N$=28 (Figs.~\ref{fig:N28_isotropic_single_even} and ~\ref{fig:N28_isotropic_single_odd}) the disturbance is localized. Upon decreasing the chain length to $N$=20 and $N$=10 the disturbance extents to (almost) the entire chain, which is a finite size effect.
What is observed (see Appendix~\ref{sec:app_corr}) is that the correlations $\langle S_1^\alpha S_{1+m}^\alpha \rangle $ are very localized both in the ground state and the state that results from the measurement. This is in accordance with the absence of long range order in the isotropic HH chain in the thermodynamic limit~\cite{BOGO86, PARK10}.
The qualitative features of the measurement are best observed for the $N=28$ system, where the width of the decoherence wave is relatively small compared to chain lengths $N=20$ and $N=10$.
What can be seen is that the forward and backward evolving waves of a single sublattice meet, and flip sign upon reflection. The waves traverse the ring, reflect again and the cycle repeats. These qualitative features can also be observed for the smaller chains $N=20$ and $N=10$. The decreasing oscillation period of the spin-up and spin-down islands for smaller $N$ are naturally explained by the fact that the wave is to traverse a shorter distance.
It is interesting to note the resemblance with standard antiferromagnetic spin-wave theory. In the spin-wave treatment one introduces creation and annihilation operators for each sublattice~\cite{PARK10}. What is observed in Fig.~\ref{fig:heisenberg_isotropic_single} is that indeed, each sublattice has an individual decoherence wave.

\subsection{Role of symmetry}\label{sec:subs_symm}
\begin{figure}
\includegraphics[width=0.45\textwidth]{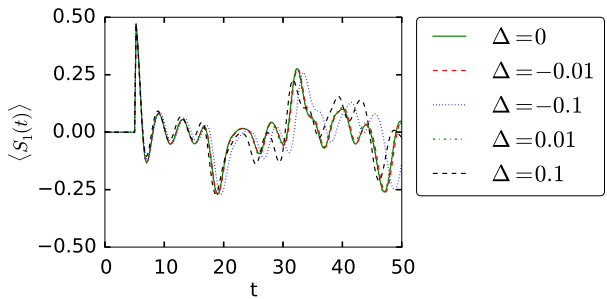}
\caption{(Color online) Time evolution of the magnetization of spin 1 in the $z$-direction ($x$-direction) for the HH with anisotropy $\Delta \geq 0$ ($\Delta <0$) and chain length $N=20$. At $t$=0 the system is prepared in the ground state and for the anisotropy $\Delta \geq 0$ ($\Delta <0$) the measurement $P^{+z}_1$ ($P^{+x}_1$) is performed at $t$=5.}
\label{fig:H_symm}
\end{figure}
In this section the importance of the system's global symmetries and its relation to the anisotropy $\Delta$ is studied.
To this end, we consider the effect of adding a small anisotropy $H^\prime$  (see Eq.~(\ref{eq:H_aniso})) to $H_0$ for both anisotropy types and examine the magnetization dynamics that results from a measurement.
According to standard terminology, the system is said to have easy plane (easy axis) magnetization 
if $\Delta < 0$ ($\Delta > 0$) ~\cite{LAND84}.

The magnetization for a chain of $N=20$ spins is presented in Fig.~\ref{fig:H_symm} in which different values of the anisotropy parameter $\Delta$ are considered.
In order to conveniently compare the different values of $\Delta$, only the magnetization of spin 1 is plotted as function of time (magnetizations of all sites are given in Appendix~\ref{sec:app_symm}).
For the anisotropies $\Delta=0.01, 0.1$ (corresponding to easy-axis magnetization) the ground state measurement is performed along the positive $z$-direction on spin 1 at $t=5$. The anisotropies $\Delta=-0.01, -0.1$ correspond to easy-plane magnetization and measurement is performed in the positive $x$-direction; the corresponding $x$-axis magnetization is depicted in the figure.
What can be seen is that for anisotropies $|\Delta| \leq 0.1$ there is some quantitative difference in the dynamics of the system. The qualitative features, however, are similar to the isotropic HH. Simulation results for chain lengths $N$ up to 28 (data not shown) indicate that this conclusion does not depend on the size of the system.

The insensitivity of the magnetization dynamics for small values of $|\Delta|$ is also suggested by considering the energy difference $\Delta E$ accompanied by the measurement. For example, the energy difference for $N$=20 is $\Delta E = 0.5936 J \, [ \, 6.667 \% \, ]$ in the absence of an anisotropy.
In this case, adding a 1 \% anisotropy (i.e. $\Delta = \pm 0.01 J$) changes the measurement induced energy difference $\Delta E$ by less than 0.04 \% relative to the ground state energy.

A priori one could think that breaking SU(2) symmetry brings about different features in the magnetization due to the reduced symmetry. For example, in earlier studies
it was suggested that for $\Delta \rightarrow 0^+$ no decoherence wave is to be observed~\cite{KATS01}.
This is shown not to be the case for $S=1/2$ (this is correct only in the limit $1/(zS) \rightarrow 0$ considered in that work). Fig.~\ref{fig:H_symm} indicates that the anisotropy $\Delta$ can not be used as a handle to align spins along a particular direction, as was assumed long ago~\cite{ANDE52}.

\subsection{Emergence of N\'eel order}\label{sec:neel_order}
\begin{figure}
\centering
\includegraphics[scale=0.45]{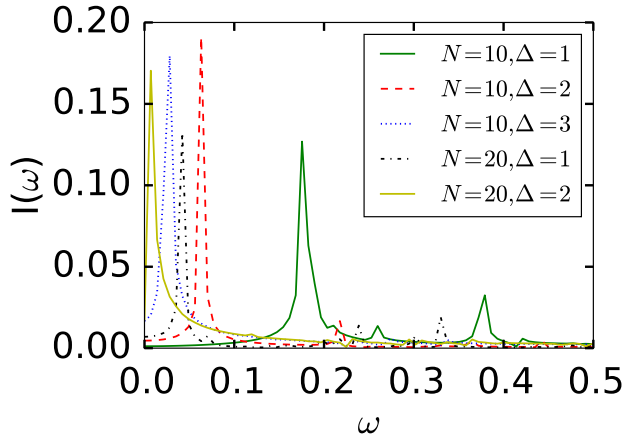}
\caption{(Color online) Fourier transform of the $z$-magnetization of spin 1 ($\langle S^z_1 (t) \rangle$) after a ground state measurement of spin 1 along the +$z$-axis. A measurement creates oscillations between sublattice configurations of which the dominant contribution decreases as function of the anisotropy $\Delta$ and chain length $N$.}
\label{fig:neel_oscillation}
\end{figure}

The ground state of $H_0$ written in the basis of local spins-up and spins-down contains a large number of components with vanishing total magnetization. As discussed in Sec.~\ref{sec:subs_symm} adding a small positive anisotropy to $H_0$ does not ensure that the ground state becomes the anti-symmetrized N\'eel state $|\psi_T\rangle = (|\psi_N\rangle - |\psi_{N^{\prime}} \rangle ) /\sqrt{2}$.
One possible way to tame the quantum fluctuations, is to increase the anisotropy $\Delta$ from zero to a large positive value. The increase of $\Delta$ relative to $J$ makes the system behave more Ising-like, and therefore increases the weight of the $|\psi_T\rangle$ contribution in the ground state.
Calculation of the ground state $|\psi_0\rangle$ of the Hamiltonian $H = H_0 + H^\prime$ for increasing anisotropy $\Delta$ indeed indicates the development of long range order (in the ground state).
The correlation function of $|\psi_0\rangle$ along the anisotropy axis increases with sites of the same sublattice and decreases for sites of inequivalent sublattices (see also Fig.~\ref{fig:ising_single_magnetization} in Appendix~\ref{sec:app_ising}).
Moreover, the norm of the correlation tends towards the maximum value of 1/4 upon increasing $\Delta$, which is characteristic for N\'eel states.
Upon performing a measurement in the +$z$-direction onto spin 1 of the ground state, the system starts to oscillate between the two sublattice configurations. This is observed in Fig.~\ref{fig:neel_oscillation} which depicts $I(\omega)$, the absolute value of the Fourier transform of $\langle S^z_1(t) \rangle$,
 as a function of the dimensionless wave number $\omega$ (that is, $\omega/(\hbar J)$).

It is seen that the frequency $\omega$ of the dominant oscillation decreases for larger values of $\Delta$.
That is, the time scale in which the state has a particular N\'eel-like configuration is increased by considering larger anisotropies.
What is more, this time scale also depends on the size of the system $N$.
The dependence on the anisotropy can be understood by considering the Ising-limit (i.e. large $\Delta$).
In this case the ground state is approximately the $|\psi_T\rangle$ state,
and measurement in the $z$-direction would fix the system to either $|\psi_N\rangle$ and $|\psi_{N^\prime} \rangle$. Hence, the frequency should go to zero as $\Delta$ becomes very large.
Similarly, the decreasing of the measurement induced energy difference $\Delta E$ (for increasing $\Delta$) can be understood in the same way. To see that, $\Delta E$ is proportional to the commutator of the projection operator with the Hamiltonian $[P_i^{\pm \alpha},H]$. Therefore, the energy difference vanishes in the Ising limit. In terms of the stability criterion as proposed in Ref.~\onlinecite{HAHN16} one might say that for larger values of $\Delta$ the system is more stable as compared to $\Delta=0$.

\subsection{Multiple measurements}\label{sec:multiple_measurement}
\begin{figure*}
\centering
\subfigure[N=10]{
\includegraphics[width=0.3\textwidth]{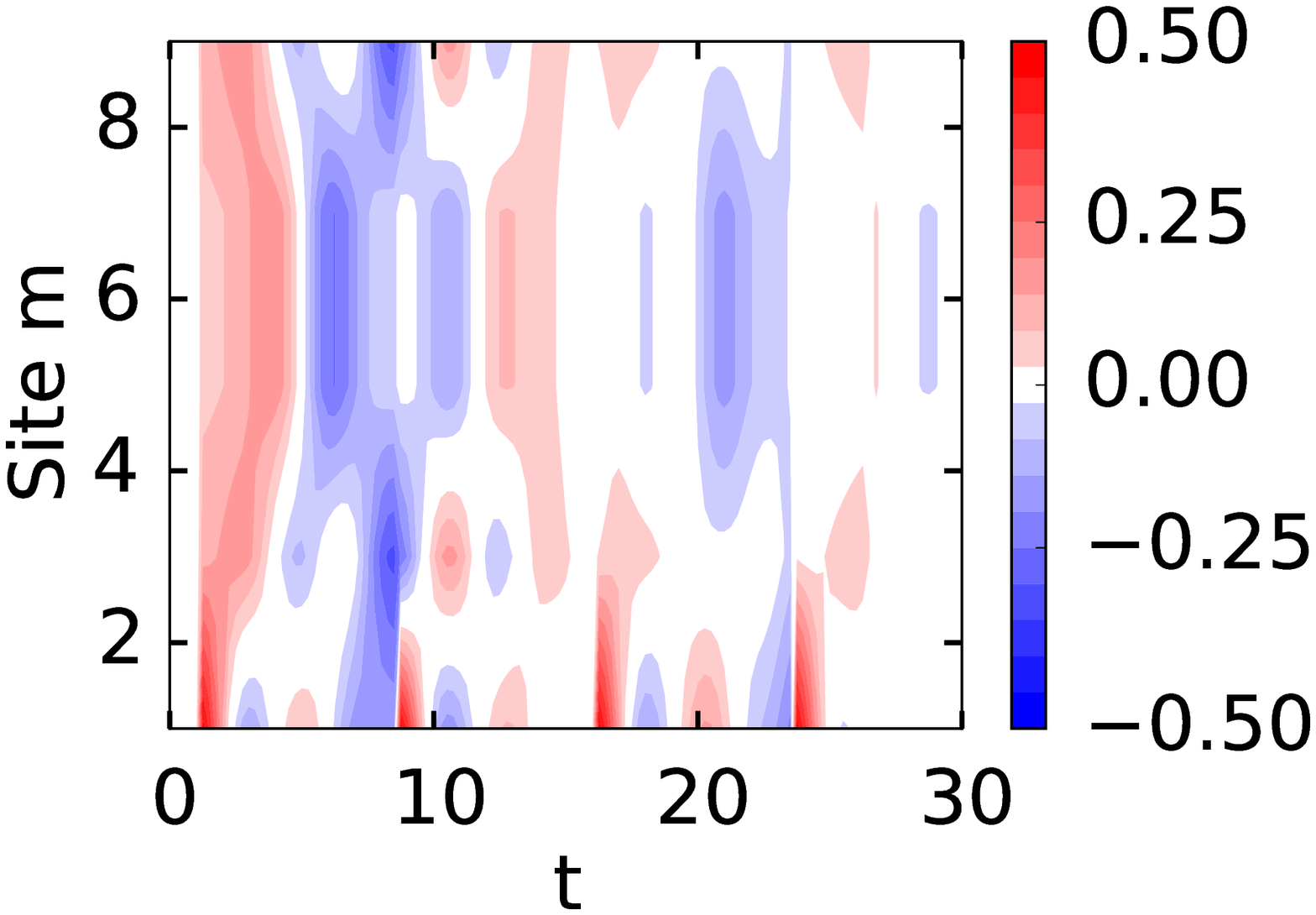} }
\subfigure[N=20]{
\includegraphics[width=0.3\textwidth]{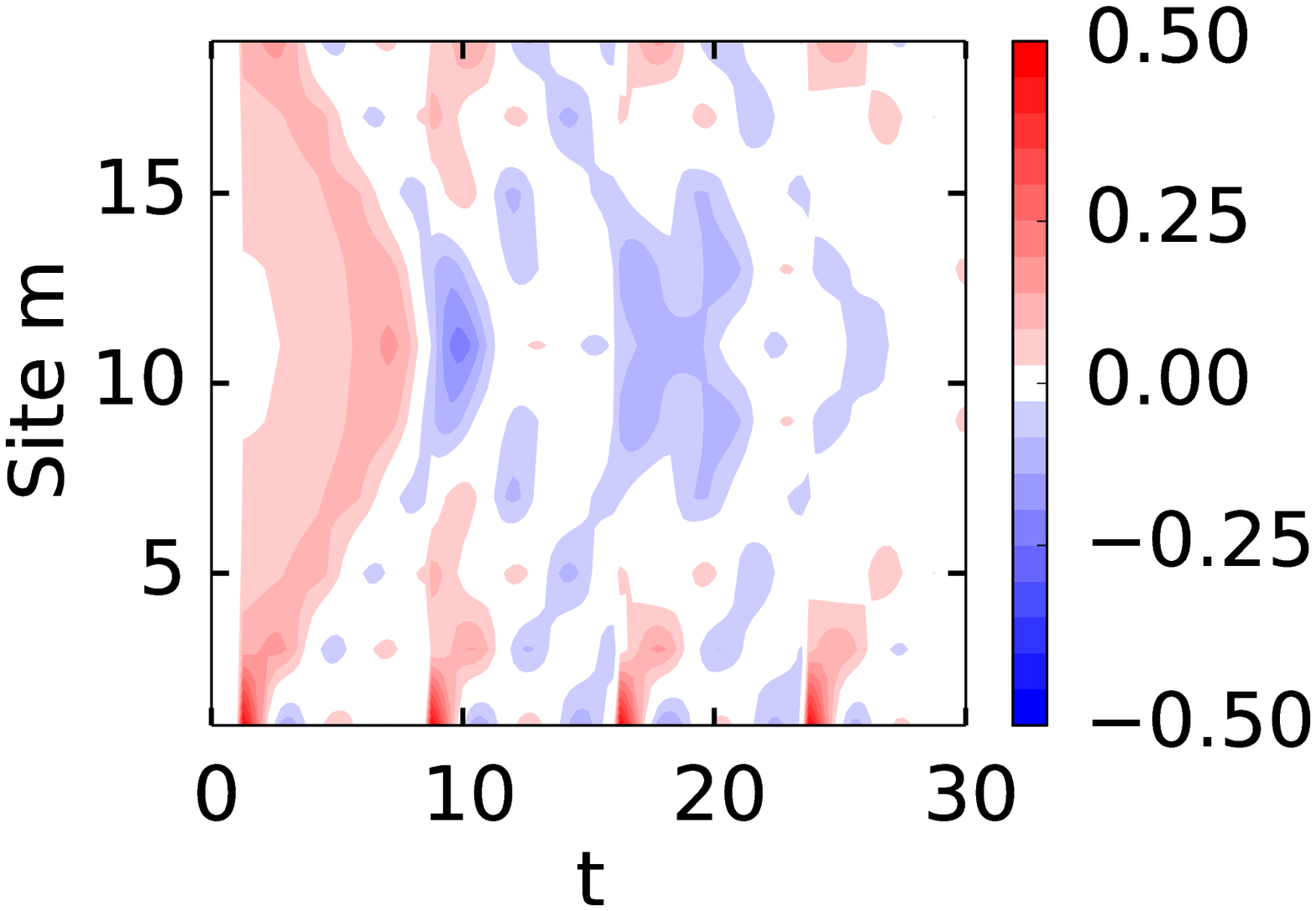}}
\subfigure[N=28]{
\includegraphics[width=0.3\textwidth]{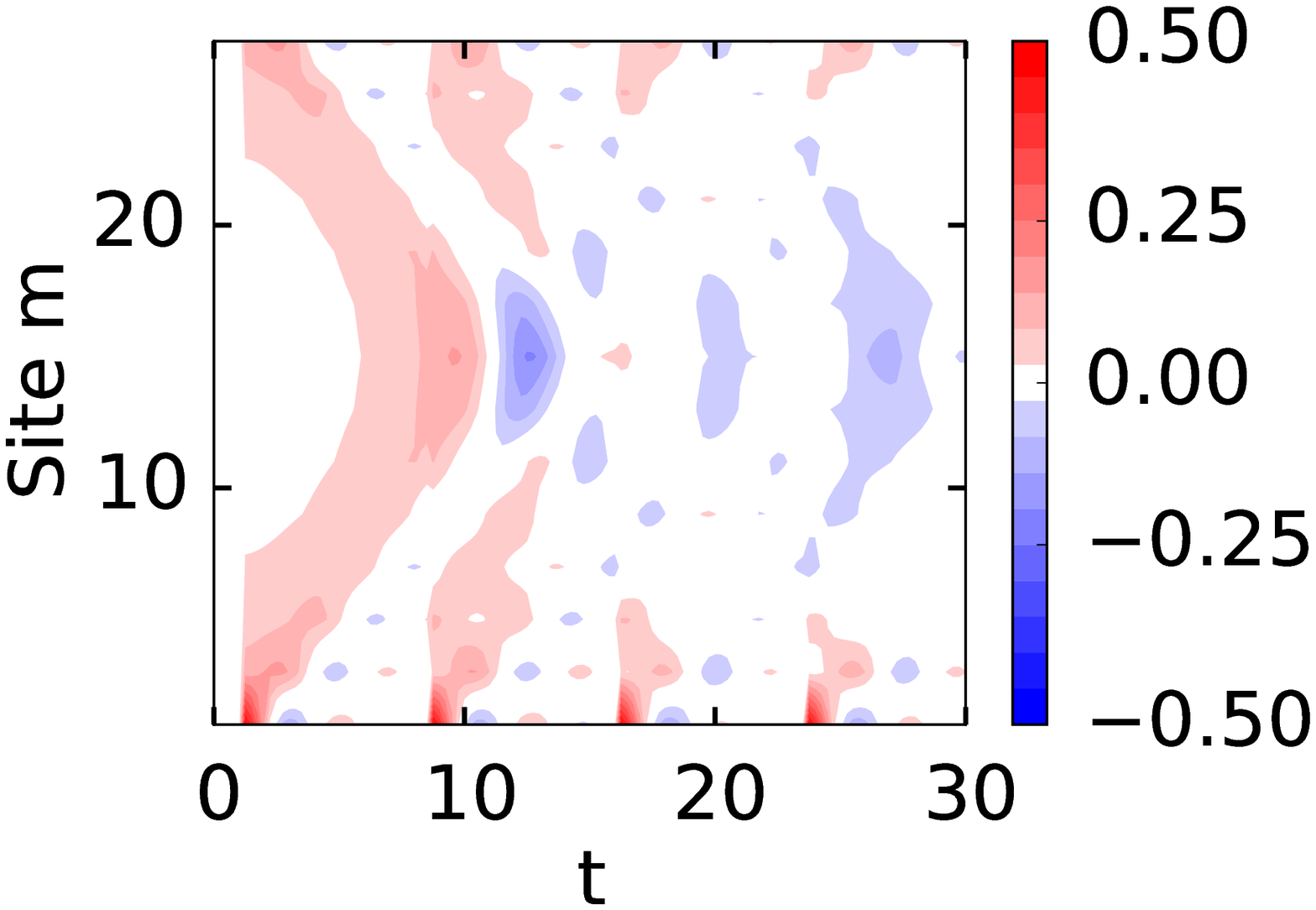}
}
\caption{(Color online) Magnetization $\langle S_m^z \rangle$ for odd values of $m$ after multiple projections on spin 1 in the $z$-direction performed at $t = 1 + 7.5m$, $m$=$0,\dots,3$ on an isotropic antiferromagnetic Heisenberg spin chain of length $N$.
}
\label{fig:heisenberg_isotropic_multiple}
\end{figure*}

A projection applied to the ground state creates a state which is entangled with numerous excitations. Indeed, a local instantaneous measurement can be interpreted as performing non-equilibrium work~\cite{YI13}. Hence, one might expect that the effects of subsequent measurements yield different dynamics.
The effect of subsequent measurements on the magnetization is addressed in Fig.~\ref{fig:heisenberg_isotropic_multiple} where the isotropic HH system is studied. This figure depicts the magnetization of the odd sites after measurement on the ground state along the $z$-axis at
$t = 1$ and subsequent measurements at $t = 1 + 7.5m$, $m=1,\dots,3$. Looking at this figure it can be seen that additional measurements do not have a pronounced effect. In particular, no disturbance waves in the magnetization are formed which resemble the waves resulting from the ground state projection.

\begin{figure}
\centering
\includegraphics[width=0.4\textwidth]{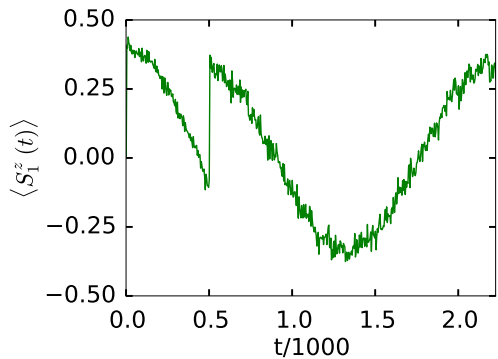}
\caption{(Color online) Magnetization $\langle S_1^z \rangle$ for $N$=20 and $\Delta=2$, projections $P_1^z$ are performed at $t=1$ and $t=500$. The subsequent measurement (at $t=500$) restores the sublattice order (close) to the state after the first measurement.}
\label{fig:seq2}
\end{figure}

The effect is somewhat different when considering the HH with additional positive anisotropy, see Fig.~\ref{fig:seq2}. A subsequent measurement temporarily restores the N\'eel-like order (the same type as described in Sec.~\ref{sec:neel_order}), after which oscillation between the two states continues.

Qualitatively the oscillation between the two N\'eel-ordered states, as observed in Figs.~\ref{fig:neel_oscillation} and~\ref{fig:seq2}, allows for an interesting interpretation.
Measurement of the ground state initially puts the system in one of the two sublattice configurations. With time evolution the state decays into a superposition, which oscillates between the two N\'eel-like states.
By performing a subsequent measurement, one resets the clock. Hence, the meta-stable state which results from measurement can be interpreted as as a manifestation of the quantum Zeno effect~\cite{MISR77, JOOS09, GRIFF05}. In the quantum Zeno analogy the undecayed state corresponds to the projected ground state and the decayed state is the state with the sublattices reversed. Subsequent measurements as described here are identical to the one considered in Ref.~\onlinecite{MISR77}, namely it is described by the operator $T_l(t) = P_l^{\pm \alpha} \exp[iHt] P_l^{\pm \alpha} $.

\section{Discussion and conclusion}
Summarizing, the effect of a localized instantaneous ground state magnetization measurement was studied by considering finite rings of antiferromagnetic spin-1/2 particles. It was found that for the isotropic HH a measurement induces a decoherence wave in each of the magnetic sublattices. Modifying the symmetry properties of the HH by introducing small anisotropies does not lead to qualitative differences. By increasing the anisotropy to the same order of magnitude as the exchange parameter, N\'eel-like order can be created by performing a measurement. With subsequent time evolution the magnetization of individual spins oscillate between the two sublattice orderings whereby additional measurements temporarily pin down a particular sublattice configuration.

The results presented here touch upon the core of quantum mechanics; namely in quantum mechanics, as opposed to classical mechanics, measurement disturbances cannot be made arbitrarily small~\cite{SCHW01}.
Indeed, the emergence of N\'eel-order due to measurement is an extreme case of such a disturbance; subsequent spin-magnetization measurements are completely determined by the outcome of the first measurement provided one performs the measurements within the Zeno-regime.

The simulation results presented here have direct experimental bearing. For an exchange value of $J$ of the order of $10^{-4}$ eV~\cite{KHAJ12} the typical time scale of the decoherence wave dynamics for $N=28$ is $10^{-11}$ s, whilst switching rates of the order of $10^8$ s$^{-1}$ have already been achieved~\cite{LOTH12}. Moreover, the time scale of the decay of N\'eel-like order depends crucially on $\Delta$ s.t. the ordering can be made stable for large time scales by tuning $\Delta$. \\

\section*{Acknowledgments}
MIK and HCD acknowledges financial support by the European Research Council, project 338957 \\FEMTO/NANO. We are grateful to Andrea Secchi for useful discussions.

\appendix

\section{Equal time correlations}\label{sec:app_corr}
\begin{figure*}
\centering
\subfigure[\, N=10]{\label{fig:N10_isotropic_corr}
\includegraphics[width=0.3\textwidth]{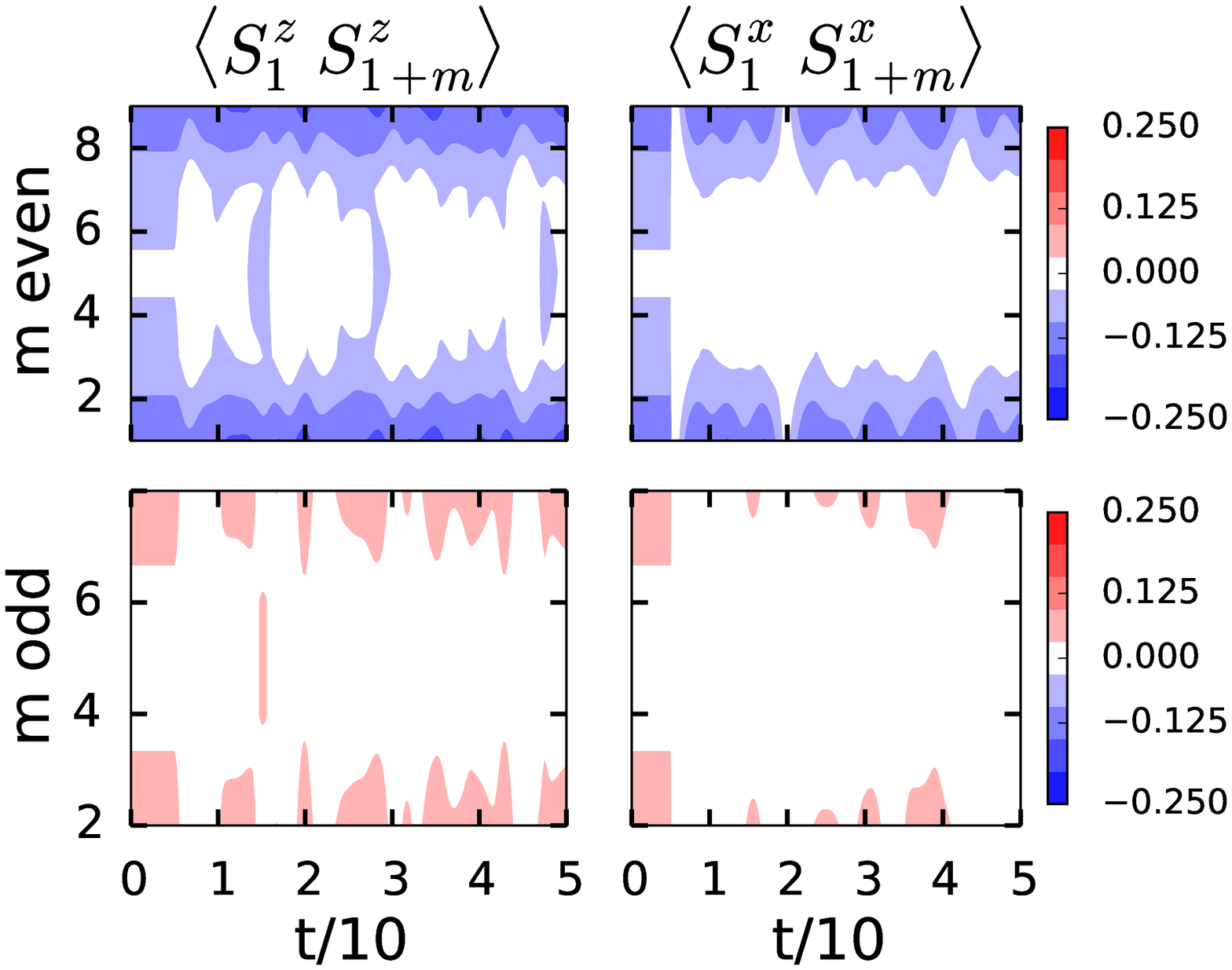} }
\subfigure[\, N=20]{\label{fig:N20_isotropic_corr}
\includegraphics[width=0.3\textwidth]{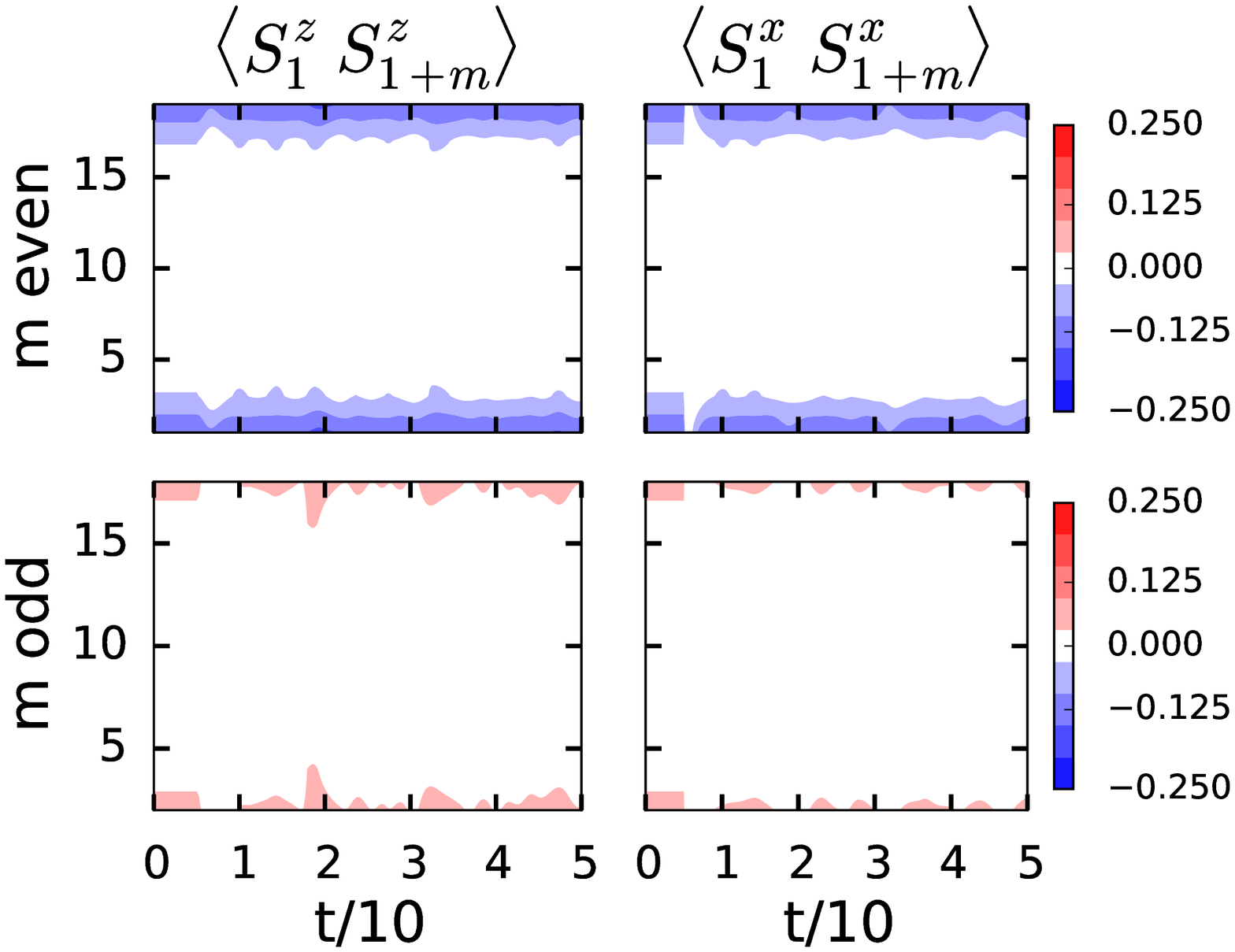} }
\subfigure[\, N=28]{\label{fig:N28_isotropic_corr}
\includegraphics[width=0.3\textwidth]{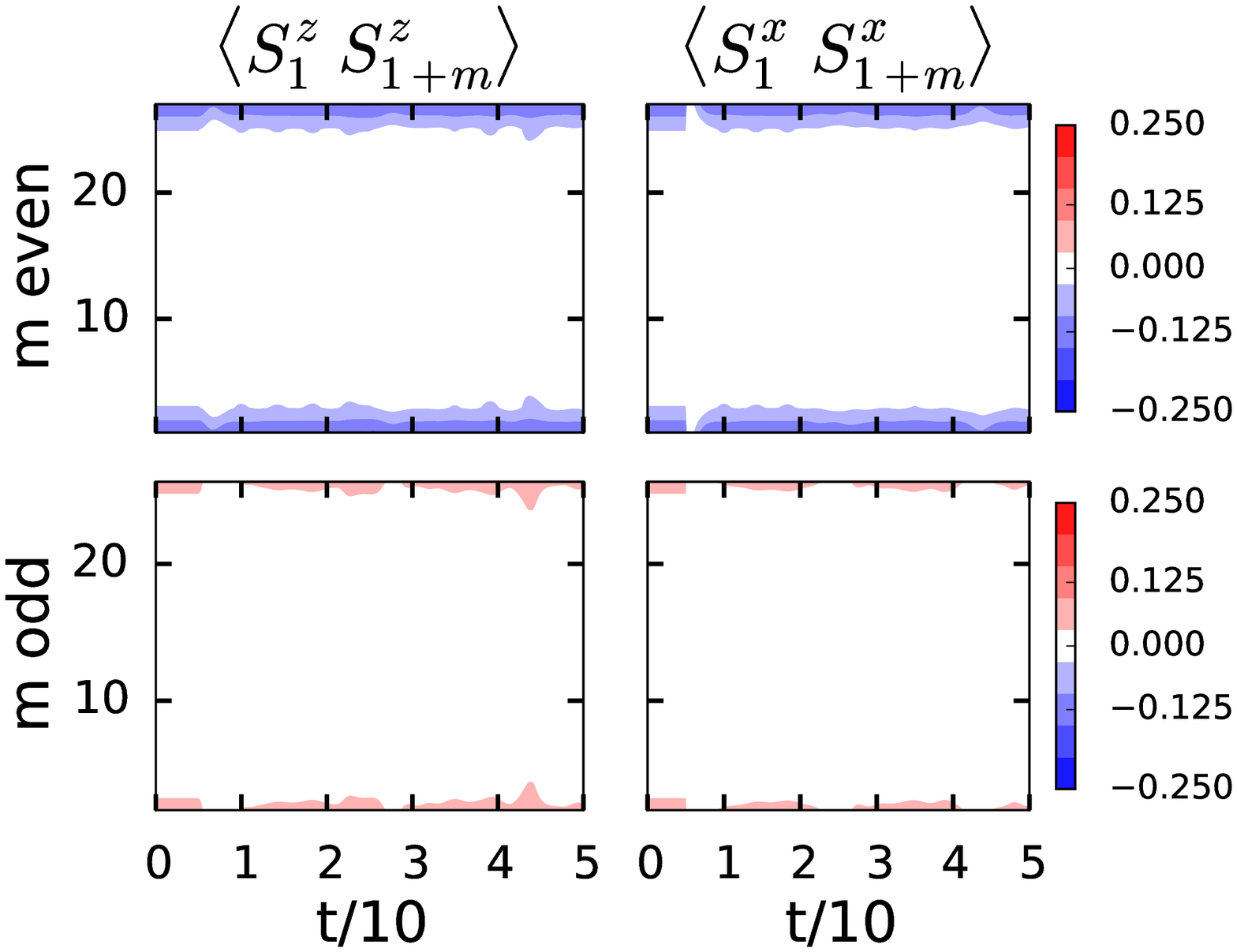} }
\caption{(Color online) The equal time correlation function $\langle S_1^\alpha S_{m+1}^\alpha \rangle$ with $\alpha={z,x}$ as a function of $m$ and the dimensionless time $t$ for the isotropic antiferromagnetic Heisenberg spin chain of length $N$ with periodic boundary conditions. The system is prepared in the ground state after which spin 1 is projected on the $+z$-axis at $t = 5$.}
\label{fig:heis_single_corr}
\end{figure*}

Subject of this Appendix are the (equal time) correlation functions $\langle S^{z}_1 S^z_{1+m} \rangle$ of the isotropic Heisenberg Hamiltonian (Eq.~(\ref{eq:HH})). In particular the system under consideration is prepared in the ground state and is subjected to a local instantaneous measurement (as described by Eq.~(\ref{eq:projector})) on spin 1 along the positive $z$-axis.
Fig.~\ref{fig:heis_single_corr} depicts the correlation function $\langle S^{z}_1 S^z_{1+m} \rangle$ and $\langle S^{x}_1 S^x_{1+m} \rangle$ (the $y$-correlations follow from symmetry) as a function of the distance $m$ and the dimensionless time $t$. Measurement is performed at $t = 5$ in the +$z$-direction and the correlations are split up in the two sublattices (corresponding to even and odd $m$) for different chain lengths $N$.
What is observed is that in the ground state correlations are short ranged. For example,  $|\langle S_1^zS_5^z\rangle| \leq 0.04$ for the three chain lengths $N=10, 20, 28$. Short ranged correlations are indeed expected considering the absence of long range order for the isotropic Heisenberg Hamiltonian (HH) in the $N \rightarrow \infty$ limit~\cite{BOGO86, PARK10}. In addition, the range of correlations is not significantly influenced by a measurement.

When looking at the $\langle S^{x}_1 S^x_{1+m} \rangle$ correlation function, it is seen that it vanishes at the instant of measurement.
This can be understood by writing the projected spin from the $z$- into the $x$-basis $|\uparrow \rangle = \left( |\leftarrow \rangle + |\rightarrow \rangle \right)/\sqrt{2}$.

The dynamics in the $\langle S^{x}_1 S^x_{1+m} \rangle$ correlation after the projection are similar to $\langle S^{z}_1 S^z_{1+m} \rangle$ in the sense that: 1) the correlations quickly decay as function of $m$ and 2) time-evolution does not radically change these characteristics.

Now consider the same isotropic HH set-up in which, after the initial ground state measurement, additional measurements are performed along the same axis (see Sec.~\ref{sec:multiple_measurement}).
It is found that no pronounced difference between the first and consecutive measurements can be observed in the correlations (figures not shown). This is to be contrasted with the magnetization (Fig.~\ref{fig:heisenberg_isotropic_multiple}), where it is no longer possible to speak of measurement induced decoherence waves.

In view of the aforementioned results for the equal time correlation functions for both single and multiple measurements,
one is led to conclude that for the ground state of the isotropic HH correlations are short ranged, and projections have little effect on this property.

\section{Ising-like system}\label{sec:app_ising}

\begin{figure*}
\centering
\subfigure[\, N=10, $\Delta=1$]{
\includegraphics[width=0.3\textwidth]{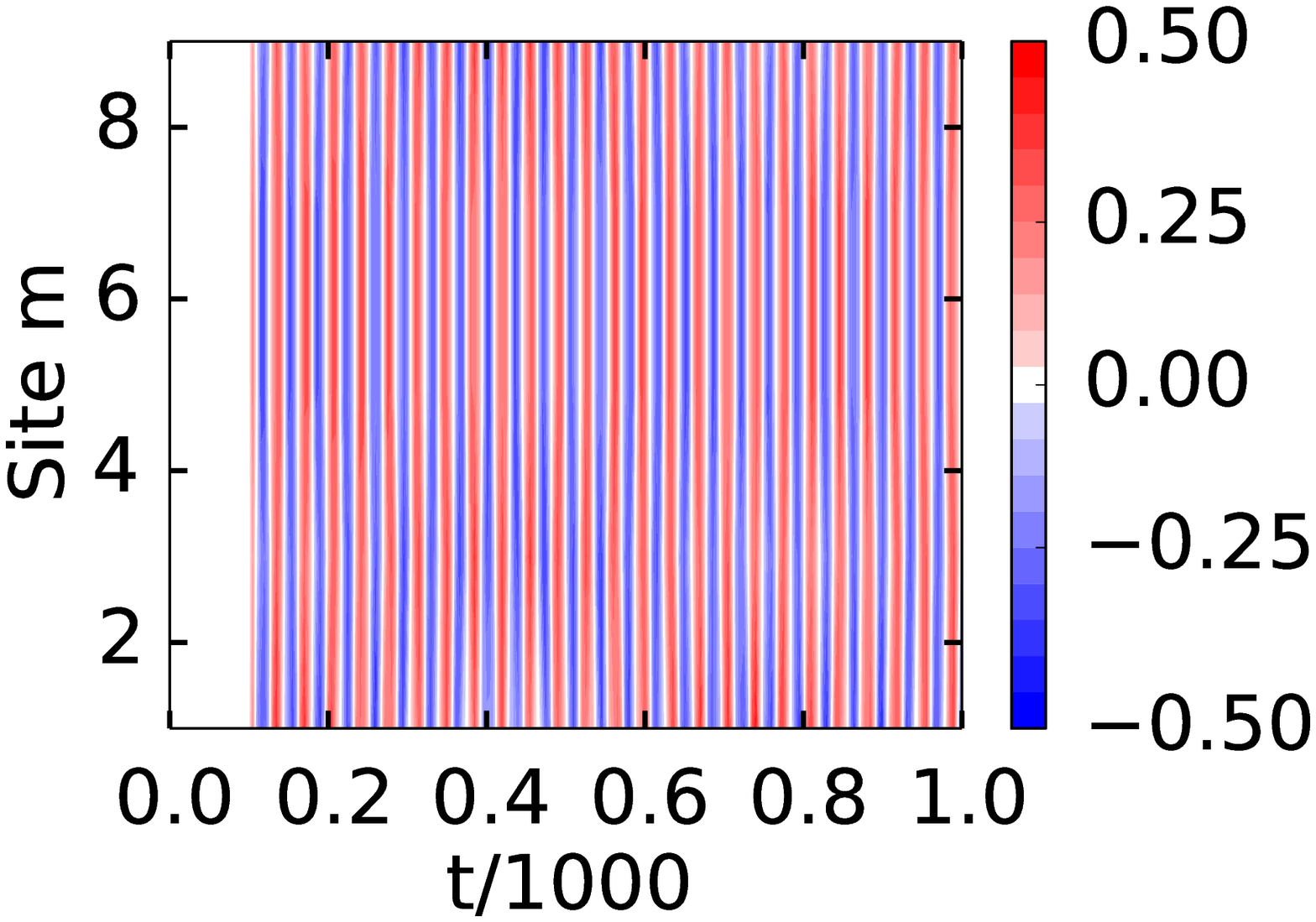} }
\subfigure[\, N=10, $\Delta=2$]{
\includegraphics[width=0.3\textwidth]{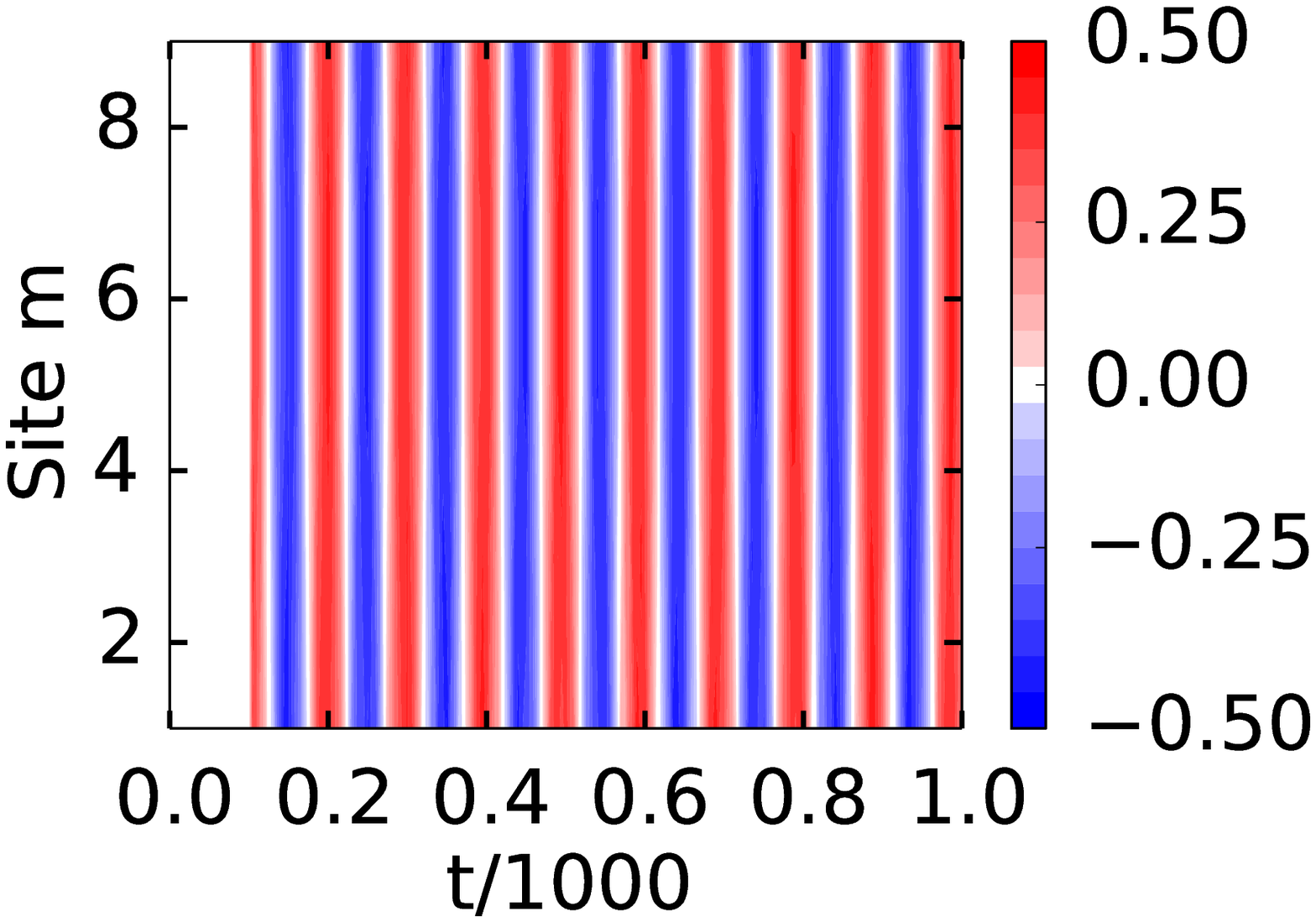}}
\subfigure[\, N=10, $\Delta=3$]{
\includegraphics[width=0.3\textwidth]{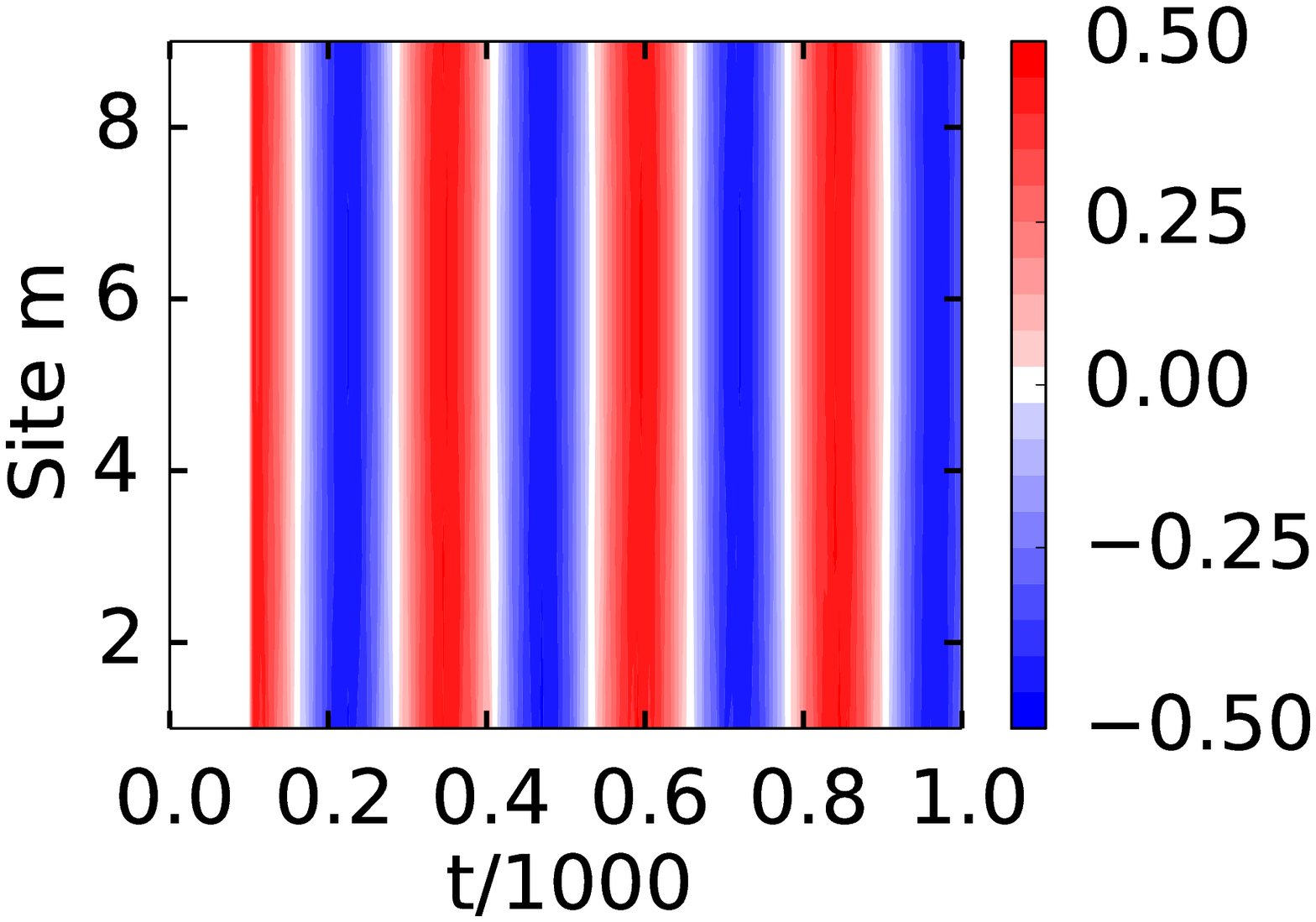} }
\subfigure[\, N=20, $\Delta=1$]{
\includegraphics[width=0.3\textwidth]{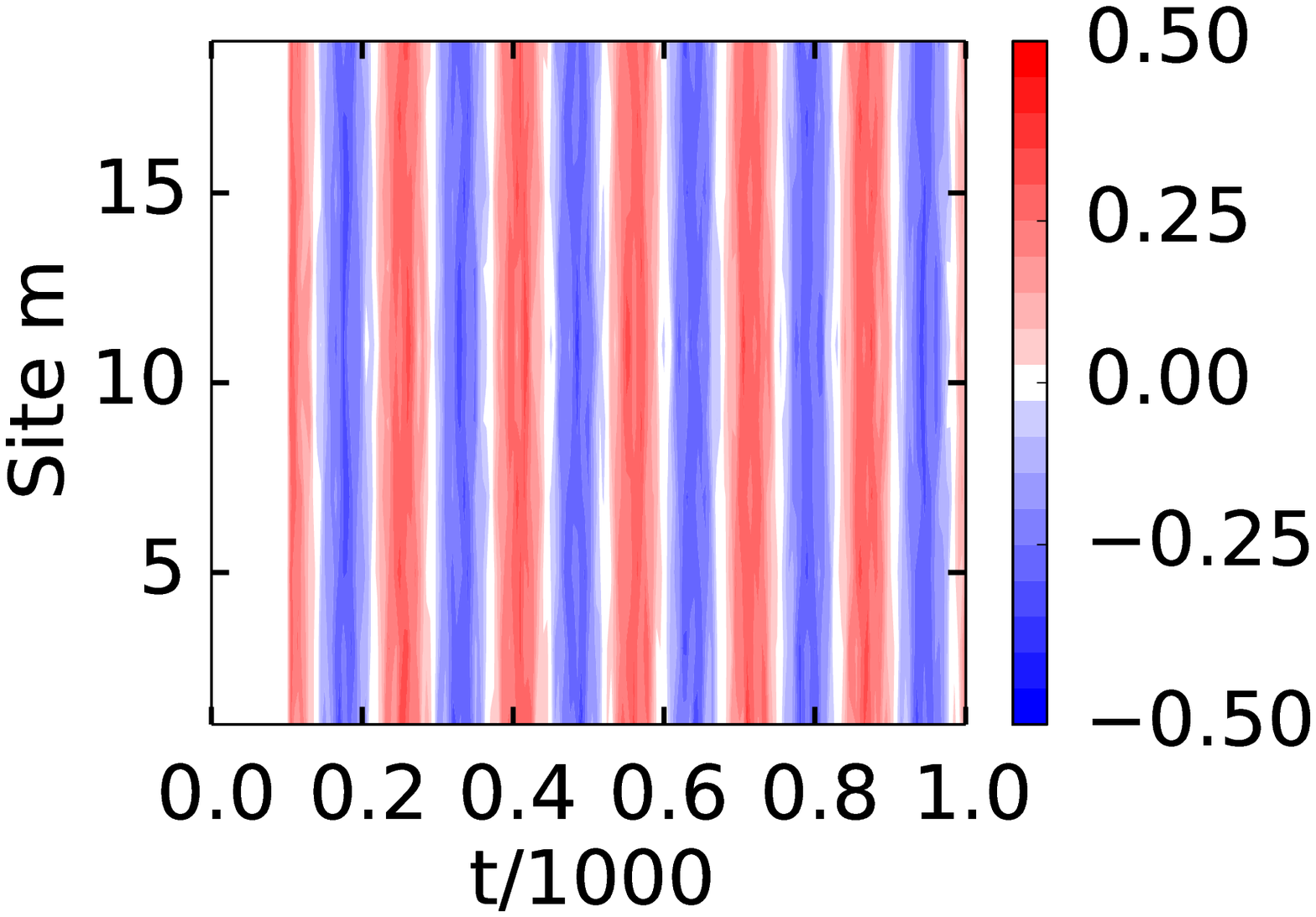} }
\subfigure[\, N=20, $\Delta=2$]{
\includegraphics[width=0.3\textwidth]{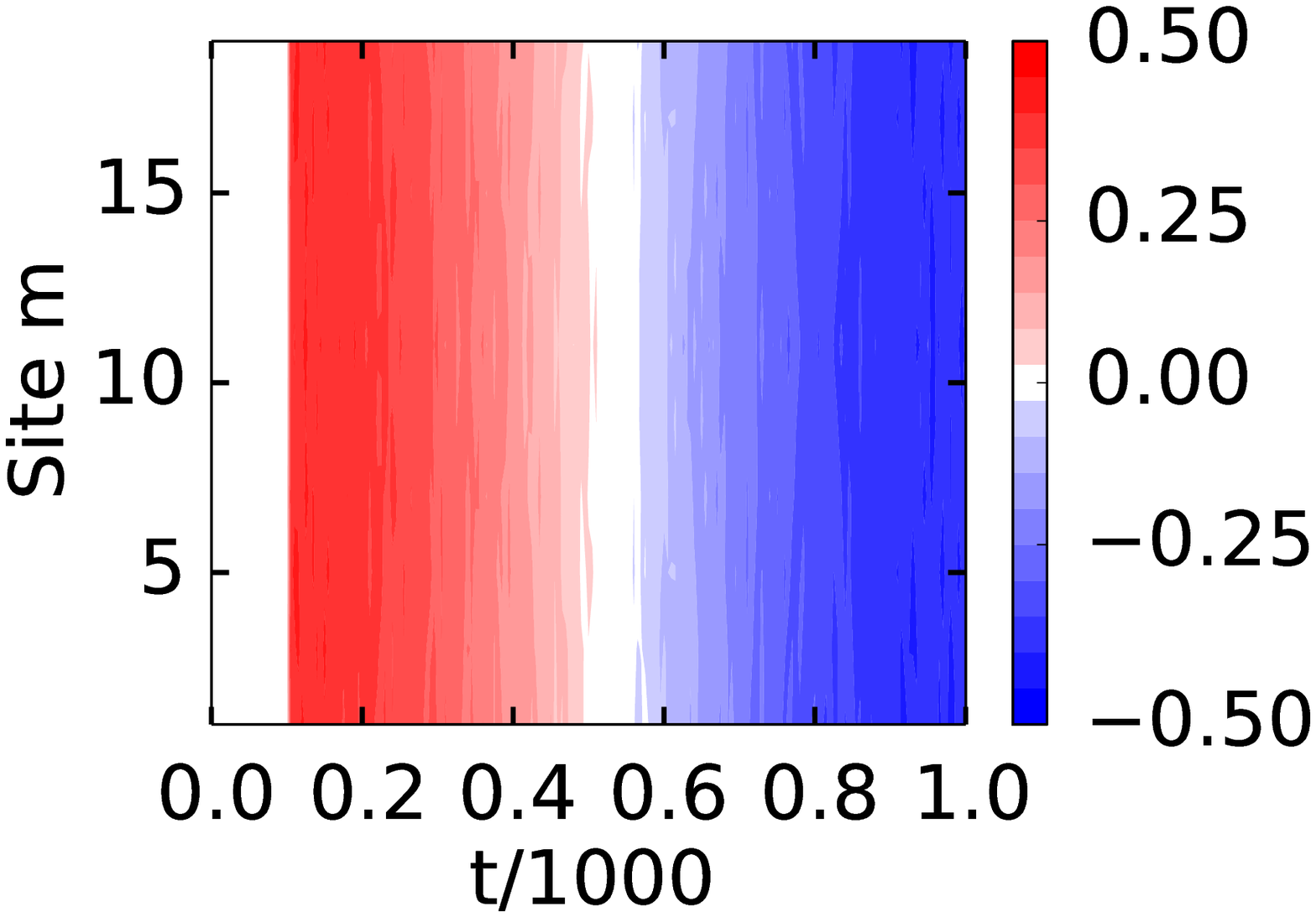}}
\subfigure[\, N=20, $\Delta=3$]{
\includegraphics[width=0.3\textwidth]{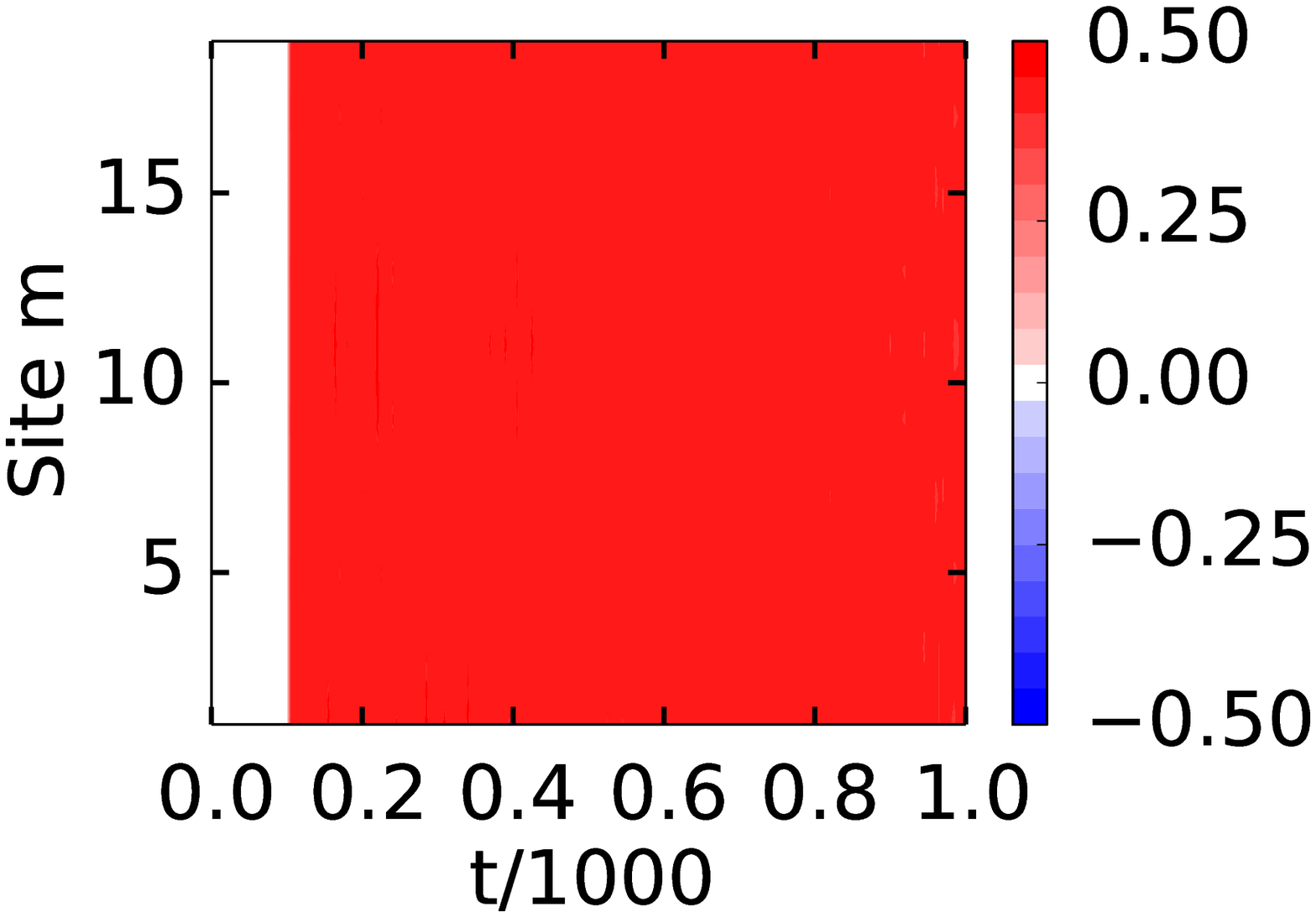} }
\caption{(Color online) Magnetization $\langle S_m^z \rangle$ for odd values of $m$ for different values of the anisotropy $\Delta$ and chain length $N$. At $t=0$ the system is prepared in the ground state, and at $t = 100$ a single measurement is performed on spin 1 along the $z$-direction.}
\label{fig:ising_single_magnetization}
\end{figure*}

\begin{figure}
\centering
\subfigure[\, $\langle S_m^z \rangle $, m odd]{
\includegraphics[width=0.22\textwidth]{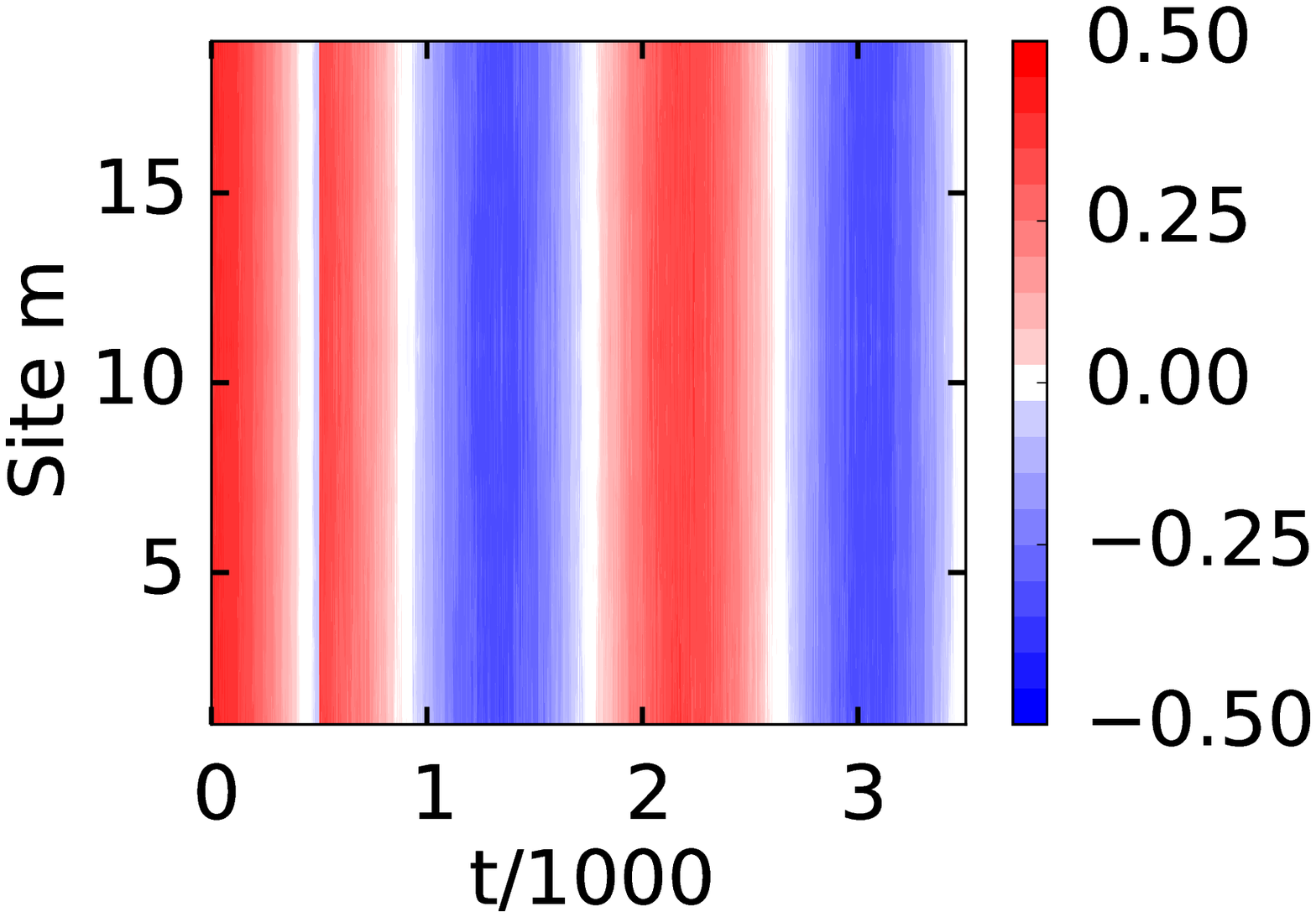} }
\subfigure[\, $\langle S_m^z \rangle $, m even]{
\includegraphics[width=0.22\textwidth]{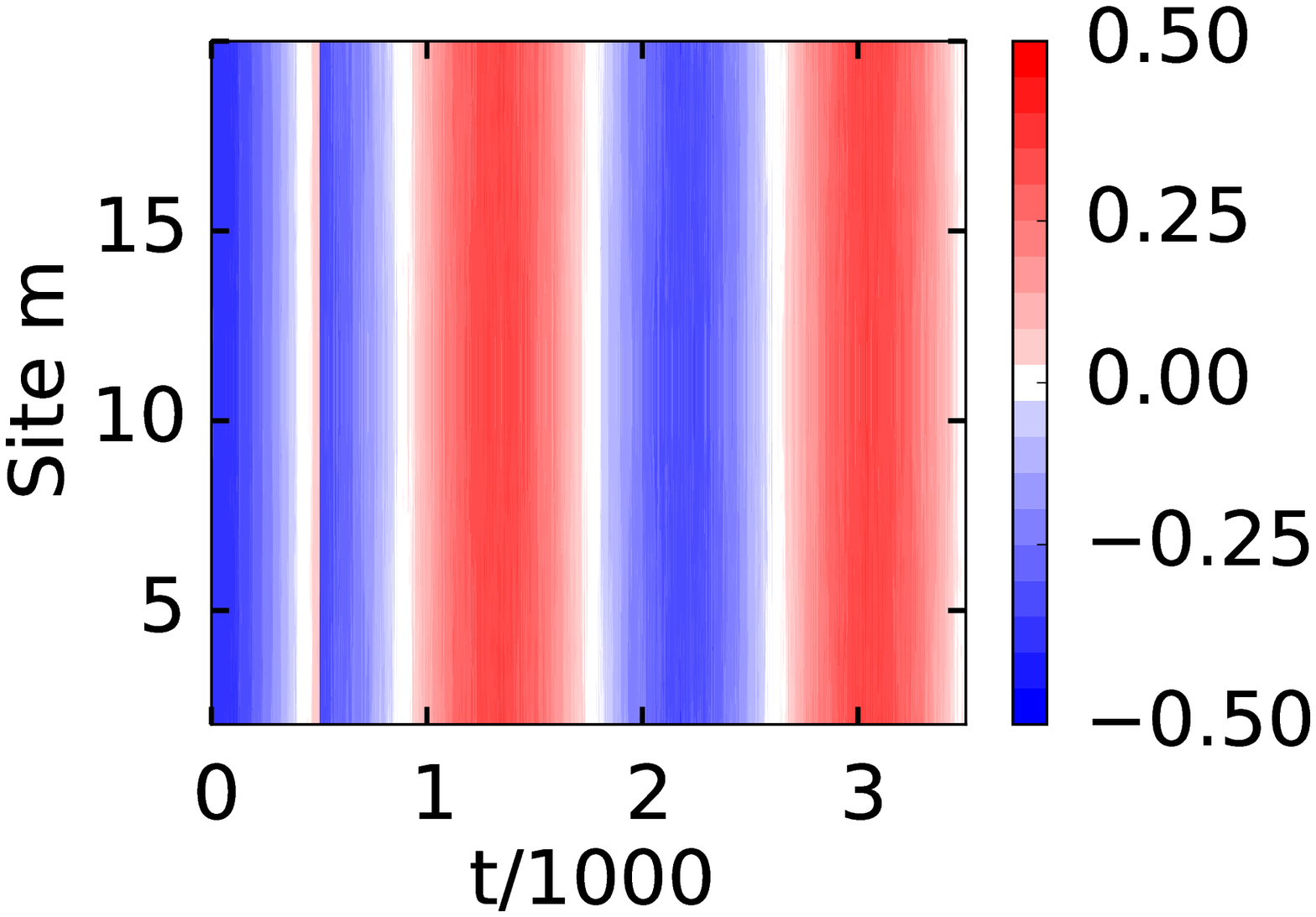} }
\caption{(Color online) Magnetization for a system with $N$=20 particles and anisotropy $\Delta=2$. The system is prepared in the ground state and two consecutive measurement are performed in the $z$-direction at $t = 1$ and $t = 500$.}
\label{fig:ising_double_magnetization}
\end{figure}

In Fig.~\ref{fig:ising_single_magnetization} the effect an easy-axis anisotropy (i.e. $\Delta > 0$) is studied, $\Delta$ being of the same order of magnitude as the exchange parameter $J$. At $t=100$ the projection $P^{+z}_1$ (parallel to the anisotropy) is applied to the ground state. What is observed is that sublattices are created as a result of measurement, the magnitude of which increases as a function of $\Delta$.
This is understood by noticing that an increase in $\Delta$ increases the weight of the N\'eel state contribution in the ground state singlet. Hence, one would expect more N\'eel-like correlations in the ground state.
The equal time correlation functions $\langle S^{\alpha}_1 S^\alpha_{1+m}\rangle$ along the axis of the anisotropy (data not shown) indicate that this is indeed the case. What is observed is that upon increasing the anisotropy $\Delta$, the ground state has increasing parallel alignment along the same sublattice.

After the von Neumann measurement, oscillation between the two sublattice configurations can be observed.
The oscillations have a well defined oscillation period (Fig.~\ref{fig:neel_oscillation}) which increases both as a function of the size of the anisotropy $\Delta$ as well as the chain length $N$.

The effect of a subsequent measurement is such that it restores the sublattice configuration to the state after the initial ground state measurement, as shown in Fig.~\ref{fig:ising_double_magnetization}. Time evolution after the second measurement shows sublattice magnetization oscillations which are analogous to the oscillation observed after the first projection.

\section{Symmetries}\label{sec:app_symm}

\begin{figure*}
\centering
\begin{tabular}{m{0.02\textwidth}m{0.05\textwidth}|*{5}{m{0.17\textwidth}}}
&& \centering{$\Delta = -0.1$} & \centering{$\Delta = -0.01$} & \centering{$\Delta = 0$} & \centering{$\Delta = 0.01$} &  \hspace{1cm} $\Delta = 0.1$\\
\hline
& N=10 & \begin{center}\includegraphics[width=0.18\textwidth]{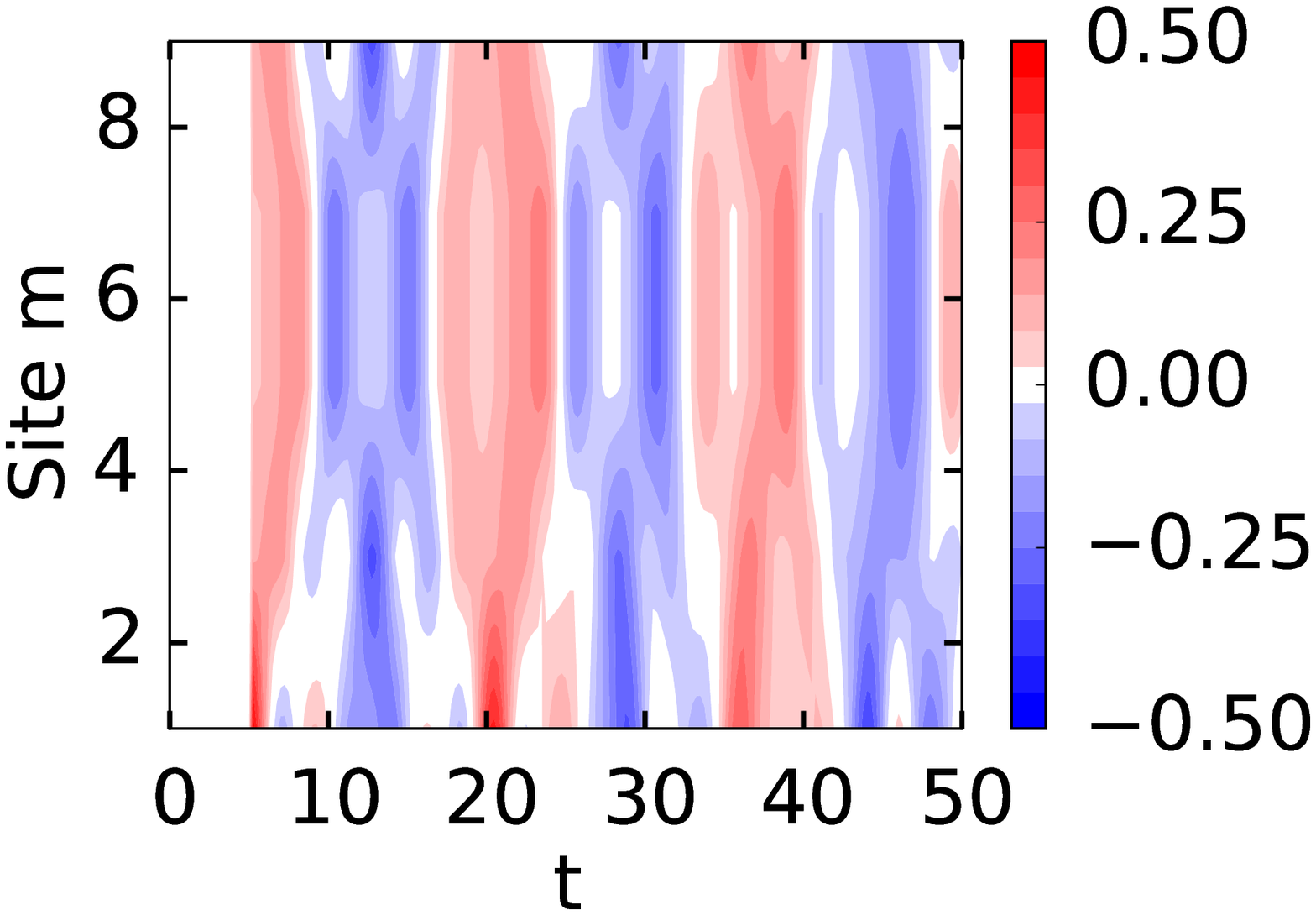}\end{center}
& \begin{center}\includegraphics[width=0.18\textwidth]{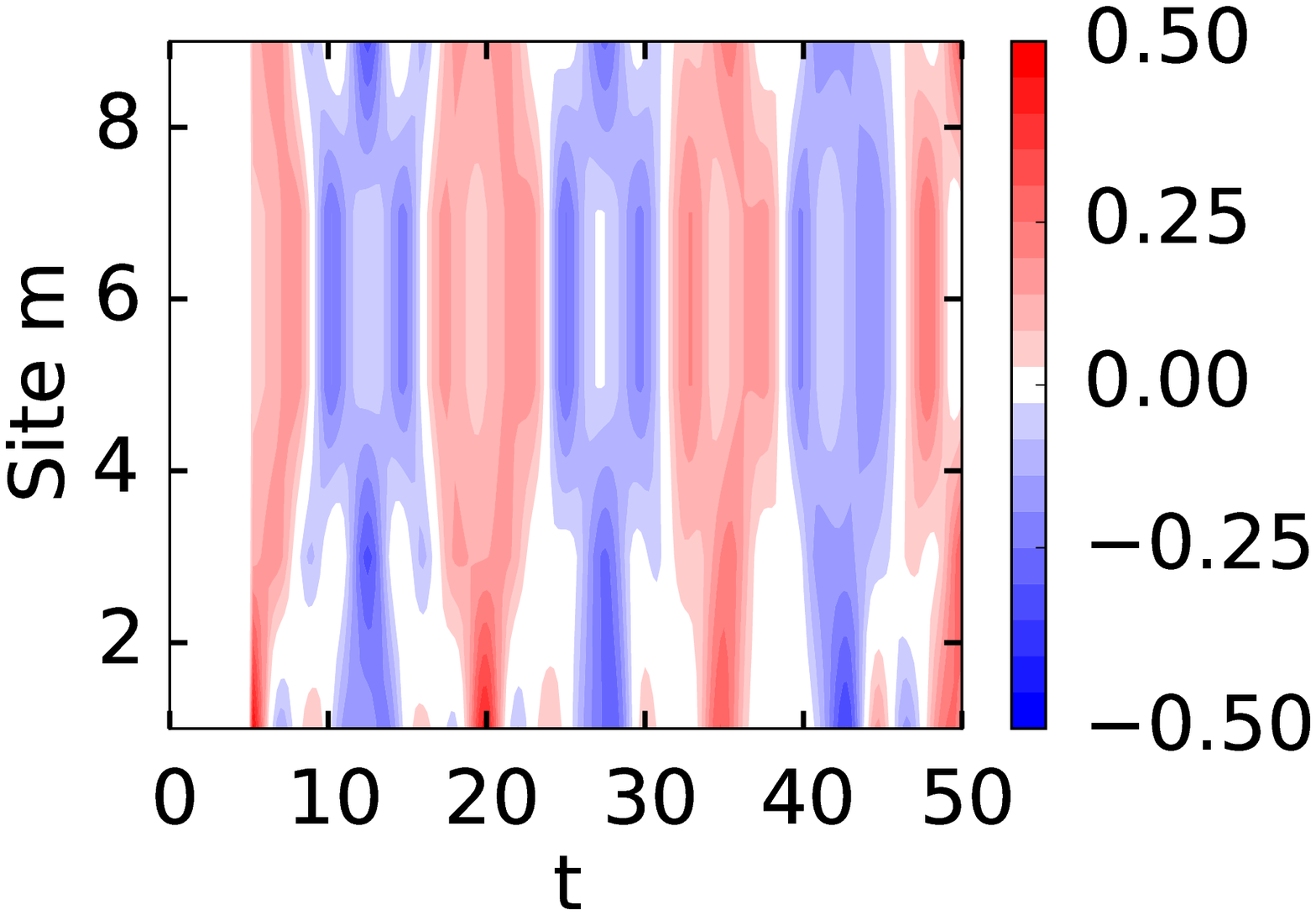}\end{center}
& \begin{center}\includegraphics[width=0.18\textwidth]{N10D0000_contour_odd_sz}\end{center}
& \begin{center}\includegraphics[width=0.18\textwidth]{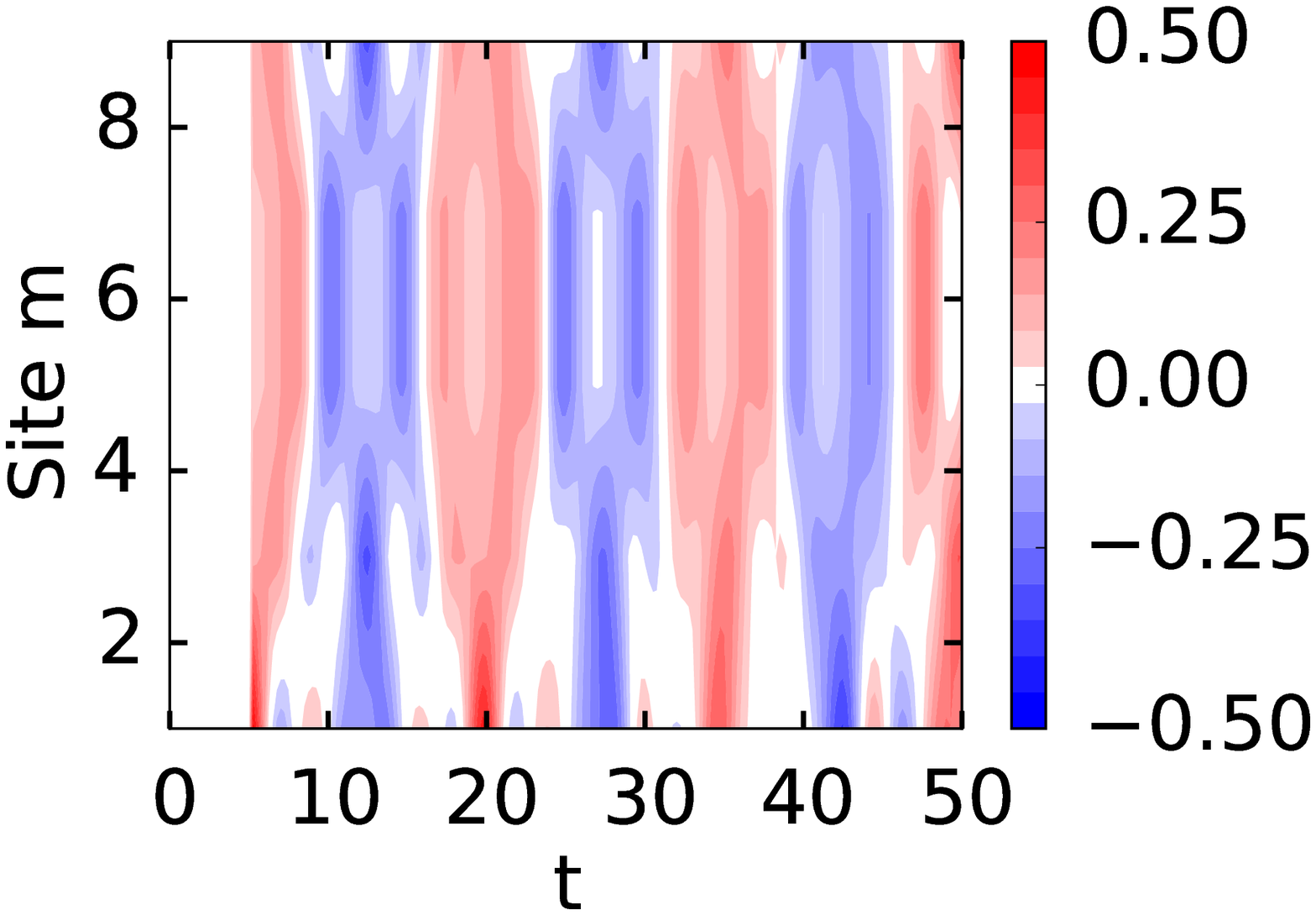}\end{center} & \begin{center}\includegraphics[width=0.18\textwidth]{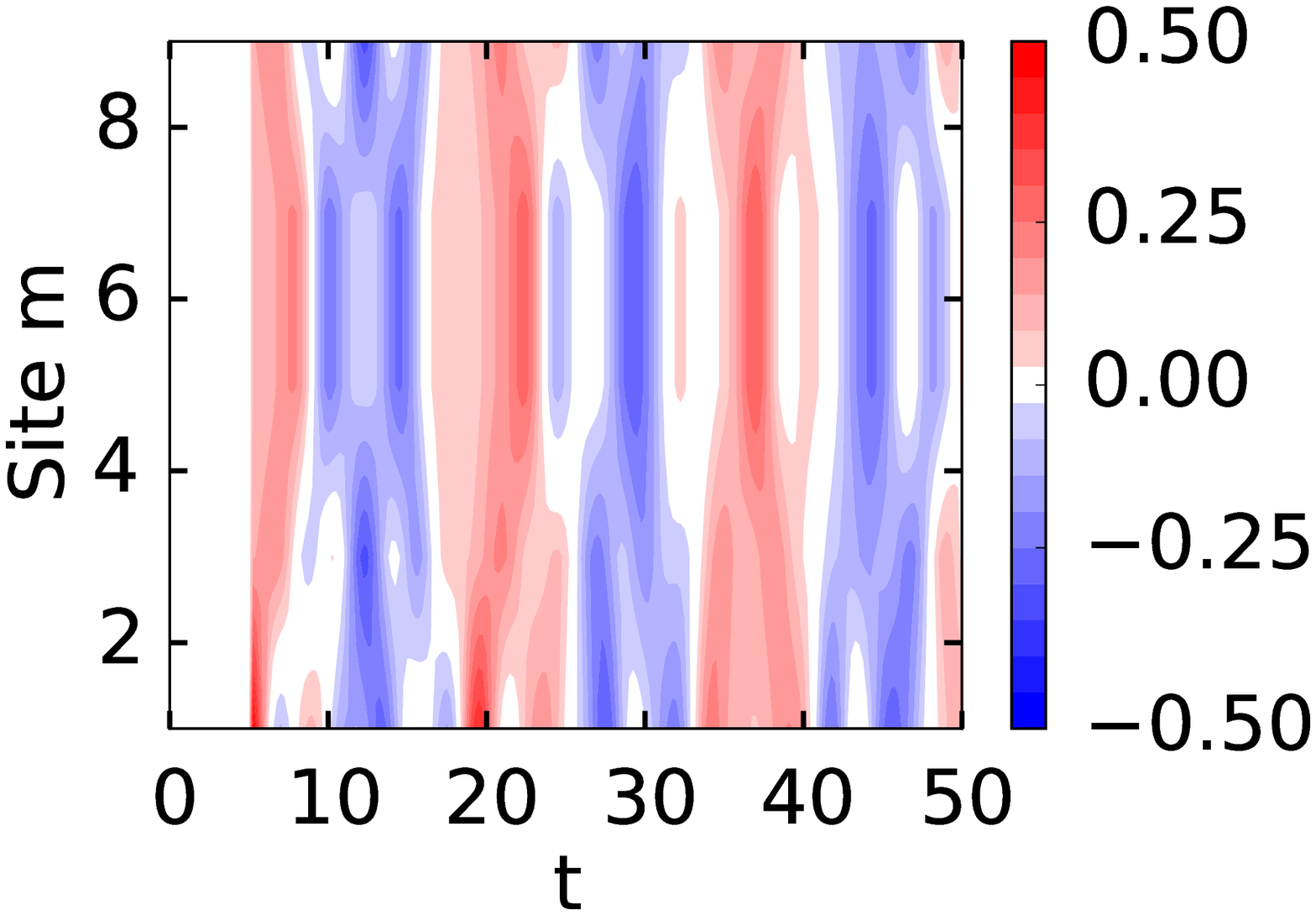}\end{center}
\\
& N=20 & \begin{center}\includegraphics[width=0.18\textwidth]{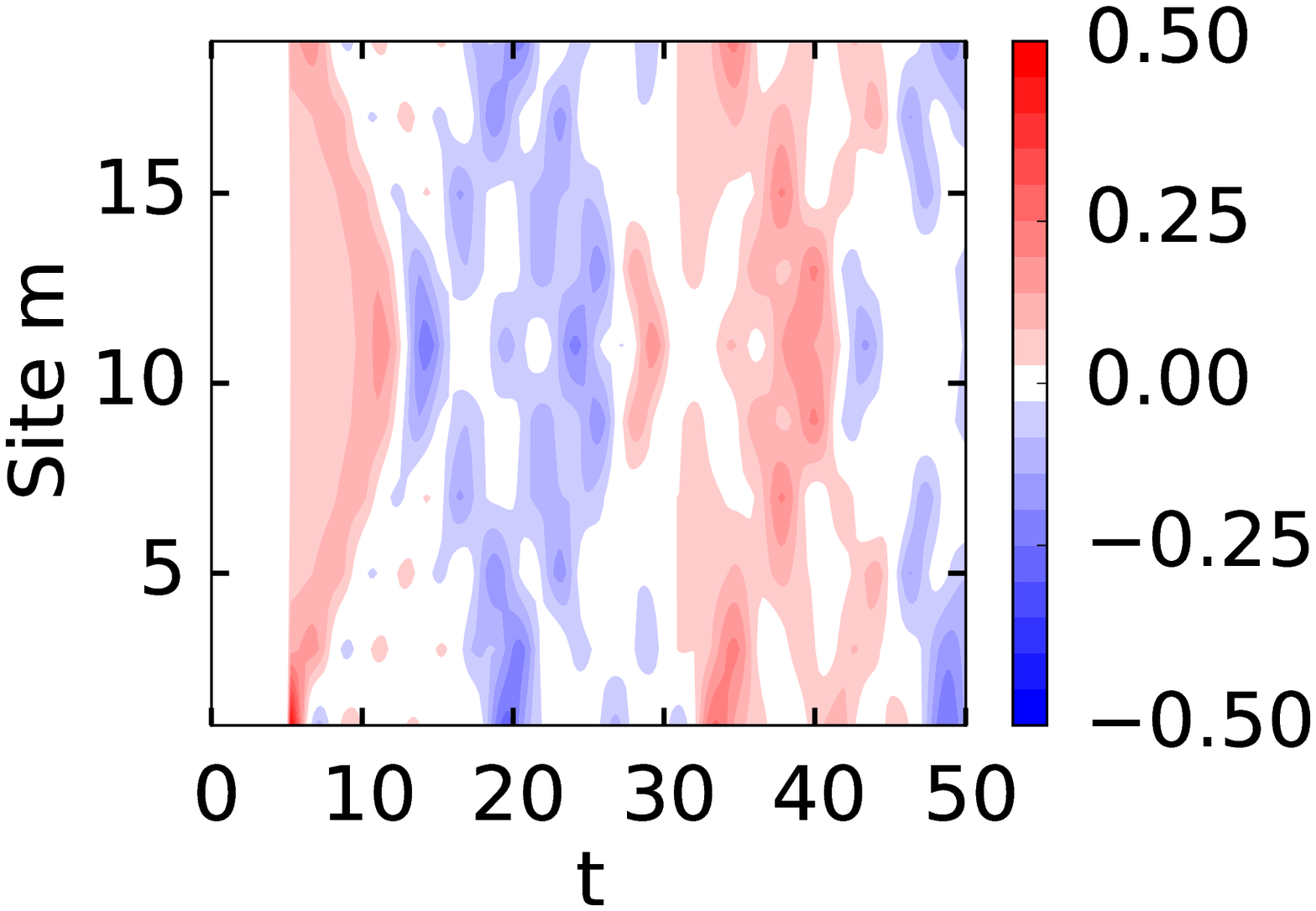}\end{center}
& \begin{center}\includegraphics[width=0.18\textwidth]{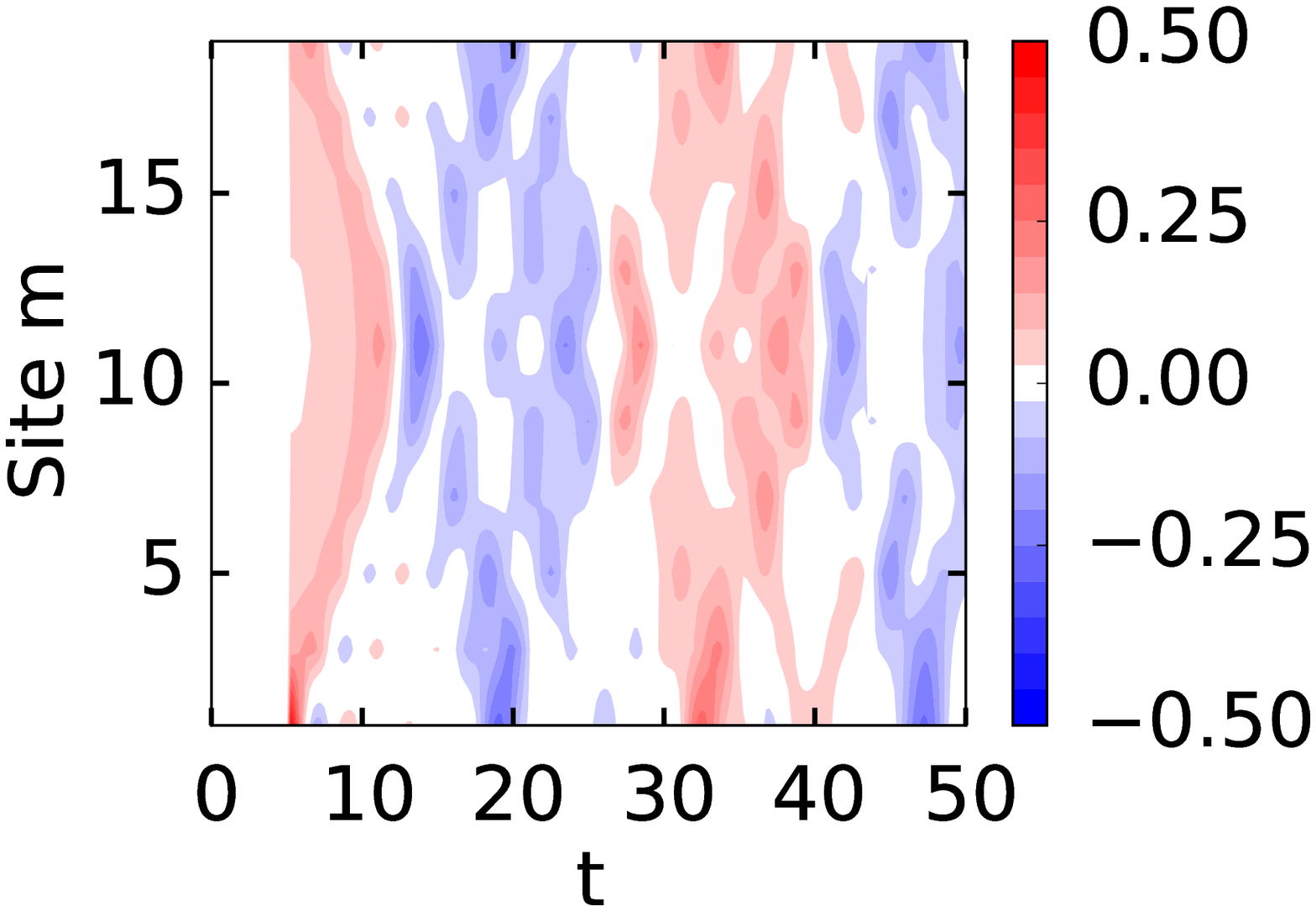} \end{center}
& \begin{center}\includegraphics[width=0.18\textwidth]{N20D0000_contour_odd_sz}\end{center}
& \begin{center}\includegraphics[width=0.18\textwidth]{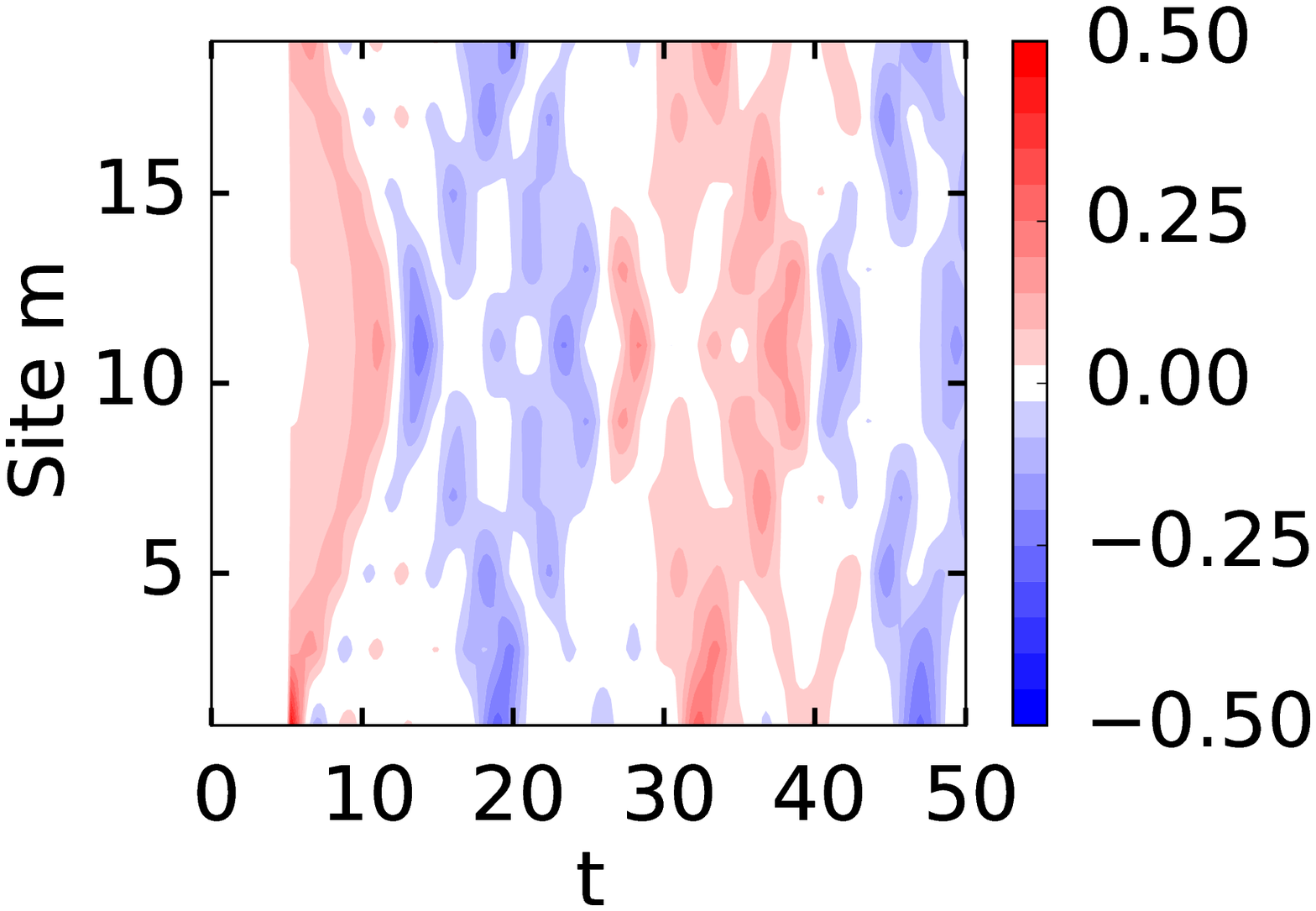}\end{center}
& \begin{center}\includegraphics[width=0.18\textwidth]{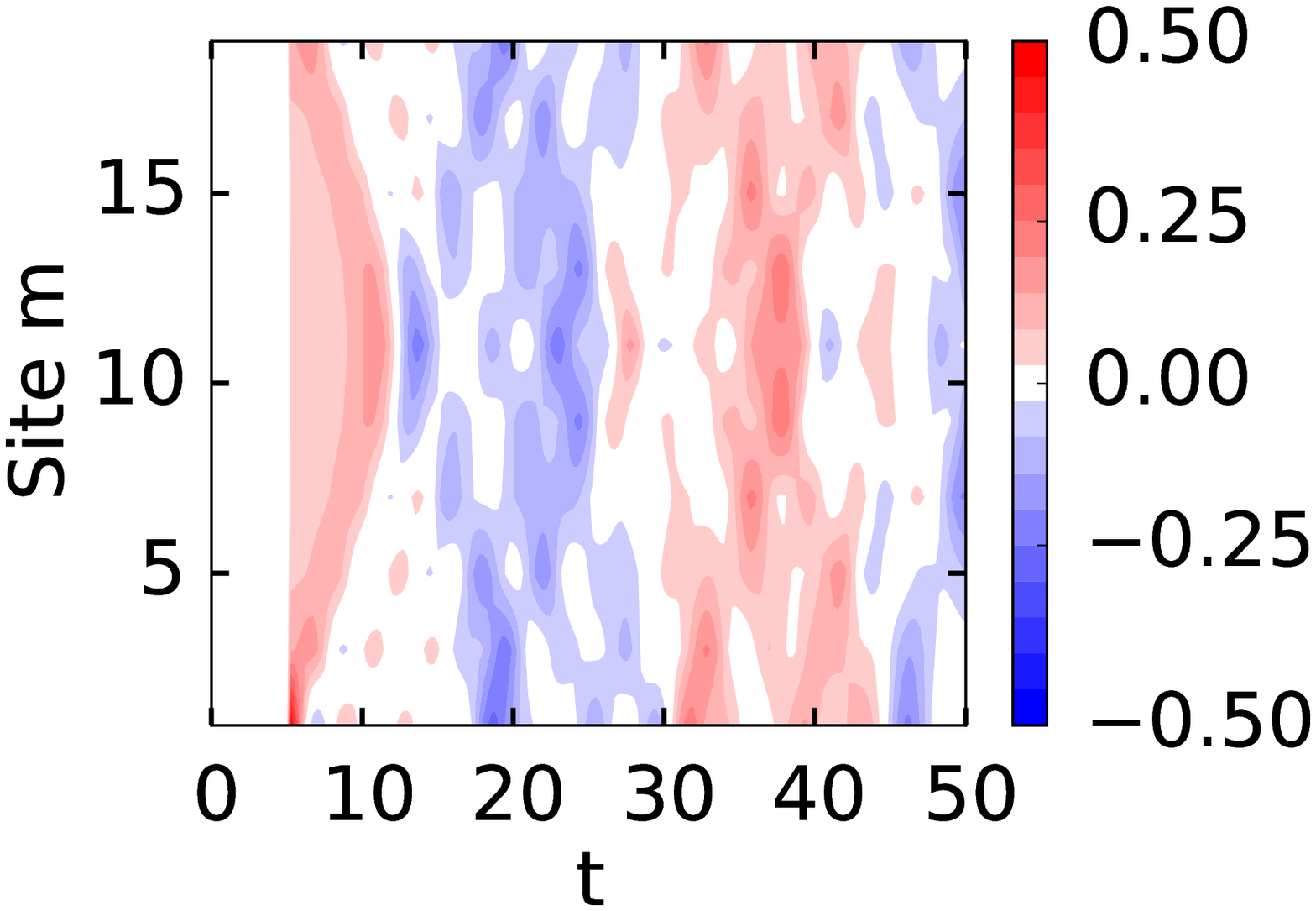}\end{center}
\\
& N=28 & \begin{center}\includegraphics[width=0.18\textwidth]{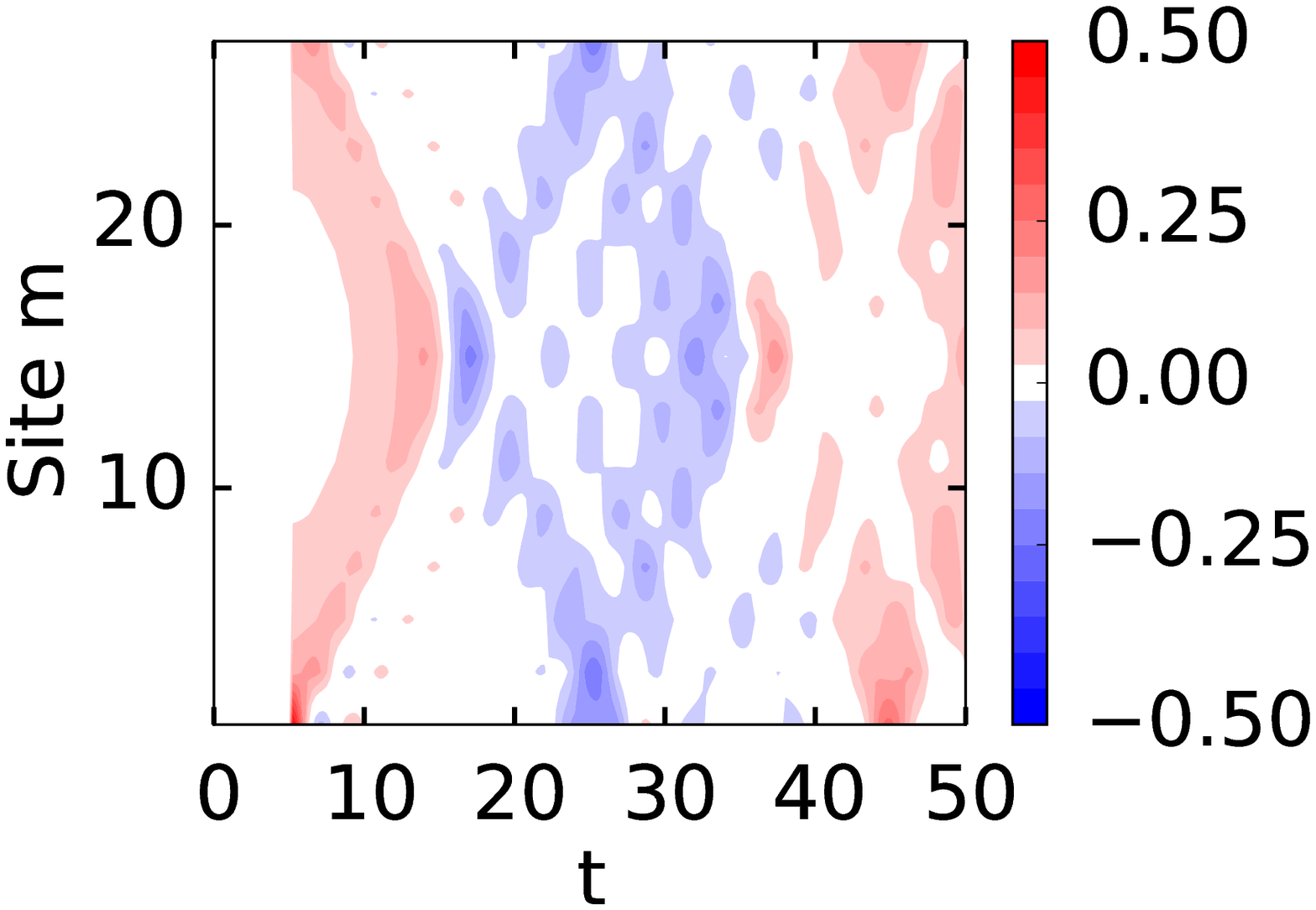}\end{center}
& \begin{center}\includegraphics[width=0.18\textwidth]{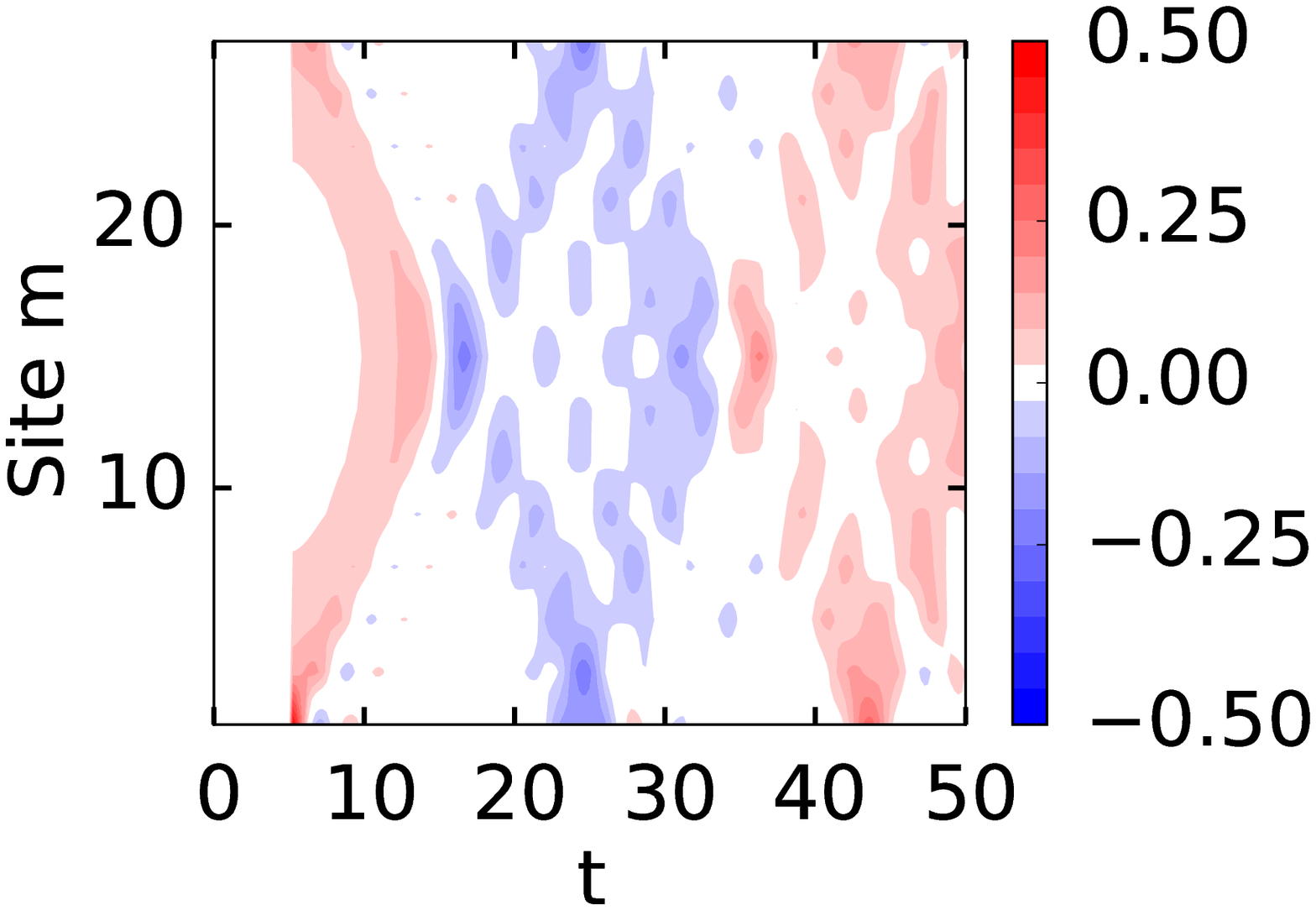}\end{center}
& \begin{center}\includegraphics[width=0.18\textwidth]{N28D0000_contour_odd_sz}\end{center}
& \begin{center}\includegraphics[width=0.18\textwidth]{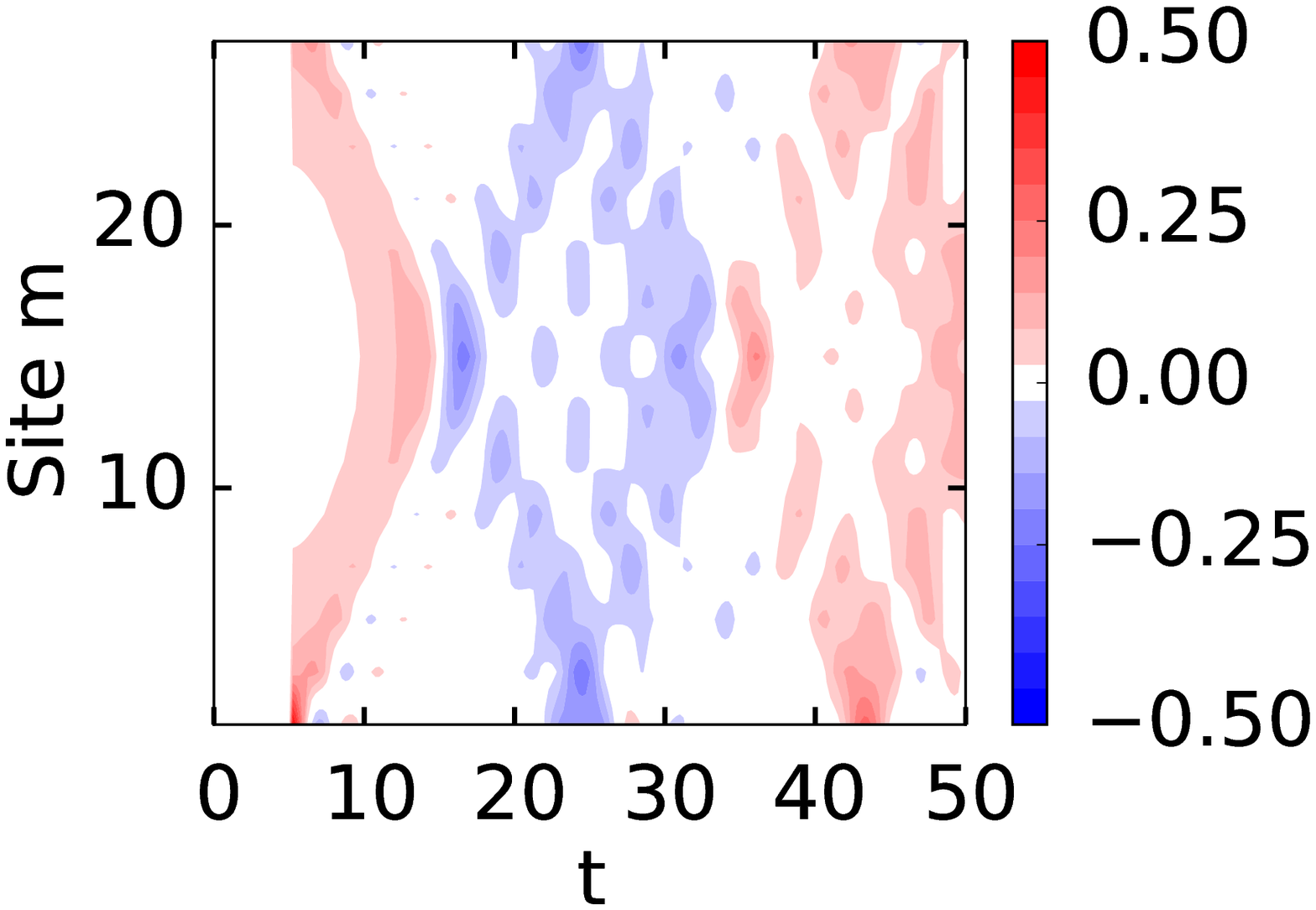}\end{center}
& \begin{center}\includegraphics[width=0.18\textwidth]{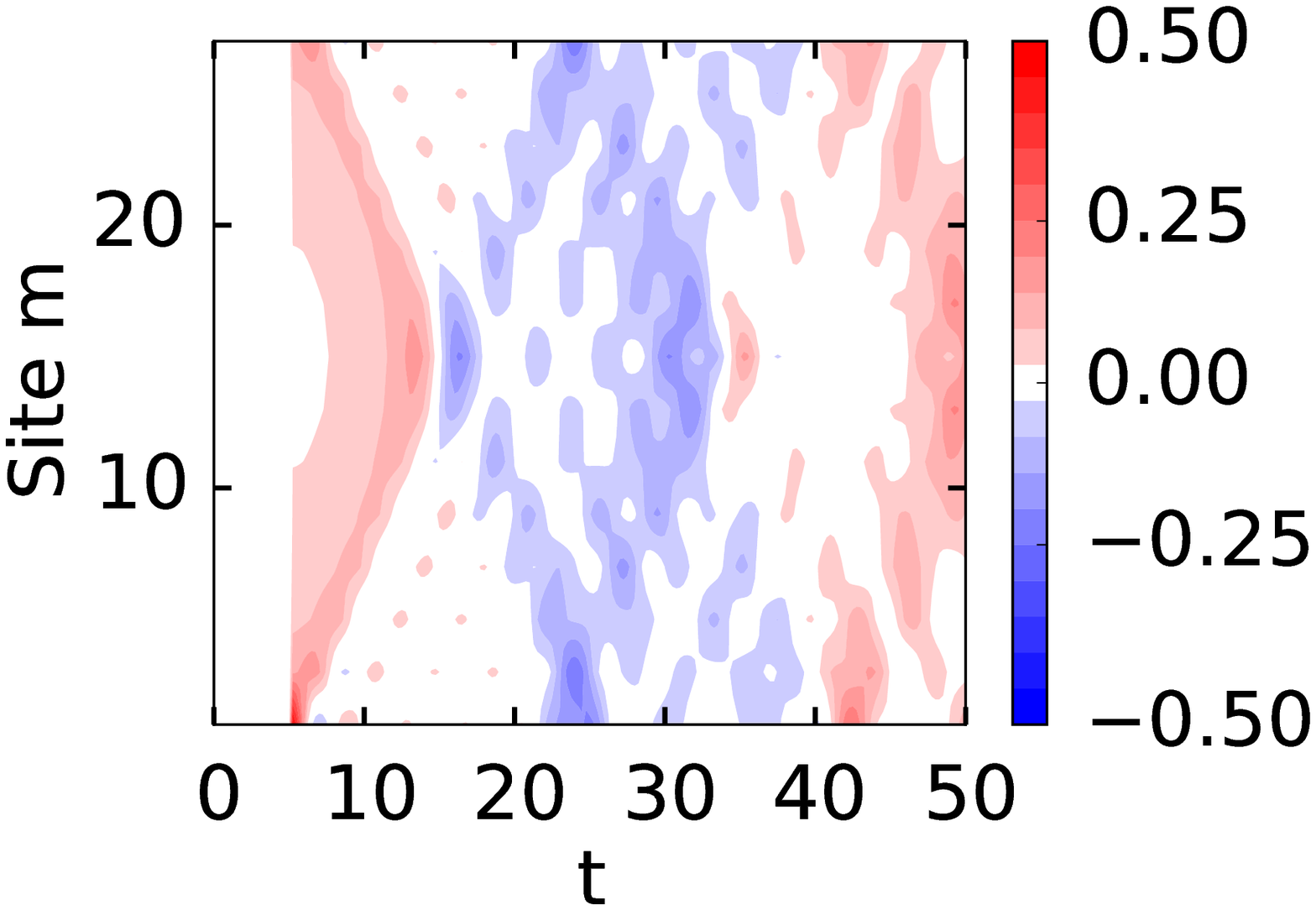}\end{center}
\\
\end{tabular}
\caption{(Color online) Magnetization $\langle S_m^\alpha \rangle$ for odd values of $m$ with $\alpha =z$ ($\alpha = x$) for anisotropy $\Delta \geq 0$ ($\Delta < 0$) for different chain lengths $N$.
The system is prepared in the ground state and a single measurement is performed on spin 1 at $t = 5$. The direction of projection is the positive $z$-axis ($x$-axis) for anisotropy $\Delta \geq 0$ ($\Delta < 0$). }

\label{fig:magnetization_symmetries}
\end{figure*}

The role of symmetry on the measurement induced dynamics can be investigated by breaking the global SU(2) symmetry.
This is done by considering an anisotropy $\Delta/J = \pm 0.1, \pm 0.01$ in the $z$-direction.
Positive (negative) anisotropy corresponds to easy-axis (easy-plane) magnetization. Therefore, measurement is performed in the $z$-direction for $\Delta >0$ and in the $x$-direction for $\Delta < 0$. The results are presented in Fig.~\ref{fig:magnetization_symmetries} where in addition to the anisotropy the chain length has been varied.
What can be noticed by looking at $\Delta=0.1$ is that the width of the decoherence wave is increased for $N$=20 and $N$=28. This can also be observed for $\Delta=-0.1$, but to a lesser extent. This is to be expected since anti-parallel alignment along the $z$-direction is slightly favoured for $\Delta >0 $ compared to the other axes. Similarly, for $\Delta<0$ anti-parallel alignment is favoured in the plane perpendicular to the $z$-axis but to a lesser degree due to rotational freedom in the plane.
The qualitative features from the decoherence wave are however, preserved upon adding a small anisotropy.

\bibliography{papers}

\begin{thebibliography}{52}%
\makeatletter
\providecommand \@ifxundefined [1]{%
 \@ifx{#1\undefined}
}%
\providecommand \@ifnum [1]{%
 \ifnum #1\expandafter \@firstoftwo
 \else \expandafter \@secondoftwo
 \fi
}%
\providecommand \@ifx [1]{%
 \ifx #1\expandafter \@firstoftwo
 \else \expandafter \@secondoftwo
 \fi
}%
\providecommand \natexlab [1]{#1}%
\providecommand \enquote  [1]{``#1''}%
\providecommand \bibnamefont  [1]{#1}%
\providecommand \bibfnamefont [1]{#1}%
\providecommand \citenamefont [1]{#1}%
\providecommand \href@noop [0]{\@secondoftwo}%
\providecommand \href [0]{\begingroup \@sanitize@url \@href}%
\providecommand \@href[1]{\@@startlink{#1}\@@href}%
\providecommand \@@href[1]{\endgroup#1\@@endlink}%
\providecommand \@sanitize@url [0]{\catcode `\\12\catcode `\$12\catcode
  `\&12\catcode `\#12\catcode `\^12\catcode `\_12\catcode `\%12\relax}%
\providecommand \@@startlink[1]{}%
\providecommand \@@endlink[0]{}%
\providecommand \url  [0]{\begingroup\@sanitize@url \@url }%
\providecommand \@url [1]{\endgroup\@href {#1}{\urlprefix }}%
\providecommand \urlprefix  [0]{URL }%
\providecommand \Eprint [0]{\href }%
\providecommand \doibase [0]{http://dx.doi.org/}%
\providecommand \selectlanguage [0]{\@gobble}%
\providecommand \bibinfo  [0]{\@secondoftwo}%
\providecommand \bibfield  [0]{\@secondoftwo}%
\providecommand \translation [1]{[#1]}%
\providecommand \BibitemOpen [0]{}%
\providecommand \bibitemStop [0]{}%
\providecommand \bibitemNoStop [0]{.\EOS\space}%
\providecommand \EOS [0]{\spacefactor3000\relax}%
\providecommand \BibitemShut  [1]{\csname bibitem#1\endcsname}%
\let\auto@bib@innerbib\@empty
\bibitem [{\citenamefont {Khajetoorians}\ \emph {et~al.}(2011)\citenamefont
  {Khajetoorians}, \citenamefont {Wiebe}, \citenamefont {Chilian},\ and\
  \citenamefont {Wiesendanger}}]{KHAJ11}%
  \BibitemOpen
  \bibfield  {author} {\bibinfo {author} {\bibfnamefont {A.~A.}\ \bibnamefont
  {Khajetoorians}}, \bibinfo {author} {\bibfnamefont {J.}~\bibnamefont
  {Wiebe}}, \bibinfo {author} {\bibfnamefont {B.}~\bibnamefont {Chilian}}, \
  and\ \bibinfo {author} {\bibfnamefont {R.}~\bibnamefont {Wiesendanger}},\
  }\href@noop {} {\bibfield  {journal} {\bibinfo  {journal} {Science}\ }\textbf
  {\bibinfo {volume} {332}},\ \bibinfo {pages} {1062 } (\bibinfo {year}
  {2011})}\BibitemShut {NoStop}%
\bibitem [{\citenamefont {Khajetoorians}\ \emph {et~al.}(2013)\citenamefont
  {Khajetoorians}, \citenamefont {Baxevanis}, \citenamefont {H\"ubner},
  \citenamefont {Schlenk}, \citenamefont {Krause}, \citenamefont {Wehling},
  \citenamefont {Lounis}, \citenamefont {Lichtenstein}, \citenamefont
  {Pfannkuche}, \citenamefont {Wiebe},\ and\ \citenamefont
  {Wiesendanger}}]{KHAJ13}%
  \BibitemOpen
  \bibfield  {author} {\bibinfo {author} {\bibfnamefont {A.~A.}\ \bibnamefont
  {Khajetoorians}}, \bibinfo {author} {\bibfnamefont {B.}~\bibnamefont
  {Baxevanis}}, \bibinfo {author} {\bibfnamefont {C.}~\bibnamefont {H\"ubner}},
  \bibinfo {author} {\bibfnamefont {T.}~\bibnamefont {Schlenk}}, \bibinfo
  {author} {\bibfnamefont {S.}~\bibnamefont {Krause}}, \bibinfo {author}
  {\bibfnamefont {T.~O.}\ \bibnamefont {Wehling}}, \bibinfo {author}
  {\bibfnamefont {S.}~\bibnamefont {Lounis}}, \bibinfo {author} {\bibfnamefont
  {A.}~\bibnamefont {Lichtenstein}}, \bibinfo {author} {\bibfnamefont
  {D.}~\bibnamefont {Pfannkuche}}, \bibinfo {author} {\bibfnamefont
  {J.}~\bibnamefont {Wiebe}}, \ and\ \bibinfo {author} {\bibfnamefont
  {R.}~\bibnamefont {Wiesendanger}},\ }\href@noop {} {\bibfield  {journal}
  {\bibinfo  {journal} {Science}\ }\textbf {\bibinfo {volume} {339}},\ \bibinfo
  {pages} {55 } (\bibinfo {year} {2013})}\BibitemShut {NoStop}%
\bibitem [{\citenamefont {Vonsovsky}(1974)}]{VONS74}%
  \BibitemOpen
  \bibfield  {author} {\bibinfo {author} {\bibfnamefont {S.~V.}\ \bibnamefont
  {Vonsovsky}},\ }\href@noop {} {\emph {\bibinfo {title} {Magnetism}}},\
  Vol.~\bibinfo {volume} {2}\ (\bibinfo  {publisher} {Wiley},\ \bibinfo {year}
  {1974})\BibitemShut {NoStop}%
\bibitem [{\citenamefont {Blundell}(2001)}]{BLUN01}%
  \BibitemOpen
  \bibfield  {author} {\bibinfo {author} {\bibfnamefont {S.}~\bibnamefont
  {Blundell}},\ }\href@noop {} {\emph {\bibinfo {title} {Magnetism in Condensed
  Matter}}}\ (\bibinfo  {publisher} {Oxford Univ. Press},\ \bibinfo {year}
  {2001})\BibitemShut {NoStop}%
\bibitem [{\citenamefont {White}(2007)}]{WHIT07}%
  \BibitemOpen
  \bibfield  {author} {\bibinfo {author} {\bibfnamefont {R.~M.}\ \bibnamefont
  {White}},\ }\href@noop {} {\emph {\bibinfo {title} {Quantun Theory of
  Magnetism}}}\ (\bibinfo  {publisher} {Springer},\ \bibinfo {year}
  {2007})\BibitemShut {NoStop}%
\bibitem [{\citenamefont {Irkhin}\ and\ \citenamefont
  {Katsnelson}(1986)}]{IRKH86}%
  \BibitemOpen
  \bibfield  {author} {\bibinfo {author} {\bibfnamefont {V.~Y.}\ \bibnamefont
  {Irkhin}}\ and\ \bibinfo {author} {\bibfnamefont {M.~I.}\ \bibnamefont
  {Katsnelson}},\ }\href@noop {} {\bibfield  {journal} {\bibinfo  {journal} {Z.
  Phys. B}\ }\textbf {\bibinfo {volume} {62}},\ \bibinfo {pages} {201}
  (\bibinfo {year} {1986})}\BibitemShut {NoStop}%
\bibitem [{\citenamefont {Kittel}(2004)}]{KITT04}%
  \BibitemOpen
  \bibfield  {author} {\bibinfo {author} {\bibfnamefont {C.}~\bibnamefont
  {Kittel}},\ }\href@noop {} {\emph {\bibinfo {title} {Introduction to Solid
  State Physics}}}\ (\bibinfo  {publisher} {Wiley},\ \bibinfo {year}
  {2004})\BibitemShut {NoStop}%
\bibitem [{\citenamefont {Tennant}\ \emph {et~al.}(1995)\citenamefont
  {Tennant}, \citenamefont {Cowley}, \citenamefont {Nagler},\ and\
  \citenamefont {Tsvelik}}]{TENN95}%
  \BibitemOpen
  \bibfield  {author} {\bibinfo {author} {\bibfnamefont {D.~A.}\ \bibnamefont
  {Tennant}}, \bibinfo {author} {\bibfnamefont {R.~A.}\ \bibnamefont {Cowley}},
  \bibinfo {author} {\bibfnamefont {S.~E.}\ \bibnamefont {Nagler}}, \ and\
  \bibinfo {author} {\bibfnamefont {A.~M.}\ \bibnamefont {Tsvelik}},\
  }\href@noop {} {\bibfield  {journal} {\bibinfo  {journal} {Phys. Rev. B}\
  }\textbf {\bibinfo {volume} {52}},\ \bibinfo {pages} {13368} (\bibinfo {year}
  {1995})}\BibitemShut {NoStop}%
\bibitem [{\citenamefont {Kojima}\ \emph {et~al.}(1997)\citenamefont {Kojima},
  \citenamefont {Fudamoto}, \citenamefont {Larkin}, \citenamefont {Luke},
  \citenamefont {Merrin}, \citenamefont {Nachumi}, \citenamefont {Uemura},
  \citenamefont {Motoyama}, \citenamefont {Eisaki}, \citenamefont {Uchida},
  \citenamefont {Yamada}, \citenamefont {Endoh}, \citenamefont {Hosoya},
  \citenamefont {Sternlieb},\ and\ \citenamefont {Shirane}}]{KOJI97}%
  \BibitemOpen
  \bibfield  {author} {\bibinfo {author} {\bibfnamefont {K.~M.}\ \bibnamefont
  {Kojima}}, \bibinfo {author} {\bibfnamefont {Y.}~\bibnamefont {Fudamoto}},
  \bibinfo {author} {\bibfnamefont {M.}~\bibnamefont {Larkin}}, \bibinfo
  {author} {\bibfnamefont {G.~M.}\ \bibnamefont {Luke}}, \bibinfo {author}
  {\bibfnamefont {J.}~\bibnamefont {Merrin}}, \bibinfo {author} {\bibfnamefont
  {B.}~\bibnamefont {Nachumi}}, \bibinfo {author} {\bibfnamefont {Y.~J.}\
  \bibnamefont {Uemura}}, \bibinfo {author} {\bibfnamefont {N.}~\bibnamefont
  {Motoyama}}, \bibinfo {author} {\bibfnamefont {H.}~\bibnamefont {Eisaki}},
  \bibinfo {author} {\bibfnamefont {S.}~\bibnamefont {Uchida}}, \bibinfo
  {author} {\bibfnamefont {K.}~\bibnamefont {Yamada}}, \bibinfo {author}
  {\bibfnamefont {Y.}~\bibnamefont {Endoh}}, \bibinfo {author} {\bibfnamefont
  {S.}~\bibnamefont {Hosoya}}, \bibinfo {author} {\bibfnamefont {B.~J.}\
  \bibnamefont {Sternlieb}}, \ and\ \bibinfo {author} {\bibfnamefont
  {G.}~\bibnamefont {Shirane}},\ }\href@noop {} {\bibfield  {journal} {\bibinfo
   {journal} {Phys. Rev. Lett.}\ }\textbf {\bibinfo {volume} {78}},\ \bibinfo
  {pages} {1787} (\bibinfo {year} {1997})}\BibitemShut {NoStop}%
\bibitem [{\citenamefont {Brockmann}\ \emph {et~al.}(2014)\citenamefont
  {Brockmann}, \citenamefont {{De Nardis}}, \citenamefont {Wouters},\ and\
  \citenamefont {Caux}}]{BROC14}%
  \BibitemOpen
  \bibfield  {author} {\bibinfo {author} {\bibfnamefont {M.}~\bibnamefont
  {Brockmann}}, \bibinfo {author} {\bibfnamefont {J.}~\bibnamefont {{De
  Nardis}}}, \bibinfo {author} {\bibfnamefont {B.}~\bibnamefont {Wouters}}, \
  and\ \bibinfo {author} {\bibfnamefont {J.-S.}\ \bibnamefont {Caux}},\
  }\href@noop {} {\bibfield  {journal} {\bibinfo  {journal} {J. Phys. A: Math.
  Theor.}\ }\textbf {\bibinfo {volume} {47}},\ \bibinfo {pages} {145003}
  (\bibinfo {year} {2014})}\BibitemShut {NoStop}%
\bibitem [{\citenamefont {Marshall}(1955)}]{MARS55}%
  \BibitemOpen
  \bibfield  {author} {\bibinfo {author} {\bibfnamefont {W.}~\bibnamefont
  {Marshall}},\ }\href@noop {} {\bibfield  {journal} {\bibinfo  {journal}
  {Proc. R. Soc. A}\ }\textbf {\bibinfo {volume} {232}},\ \bibinfo {pages} {48}
  (\bibinfo {year} {1955})}\BibitemShut {NoStop}%
\bibitem [{\citenamefont {Lieb}\ \emph {et~al.}(1961)\citenamefont {Lieb},
  \citenamefont {Schultz},\ and\ \citenamefont {Mattis}}]{LIEB61}%
  \BibitemOpen
  \bibfield  {author} {\bibinfo {author} {\bibfnamefont {E.}~\bibnamefont
  {Lieb}}, \bibinfo {author} {\bibfnamefont {T.}~\bibnamefont {Schultz}}, \
  and\ \bibinfo {author} {\bibfnamefont {D.}~\bibnamefont {Mattis}},\
  }\href@noop {} {\bibfield  {journal} {\bibinfo  {journal} {Ann. Phys.}\
  }\textbf {\bibinfo {volume} {16}},\ \bibinfo {pages} {407} (\bibinfo {year}
  {1961})}\BibitemShut {NoStop}%
\bibitem [{\citenamefont {Lieb}\ and\ \citenamefont {Mattis}(1962)}]{LIEB62}%
  \BibitemOpen
  \bibfield  {author} {\bibinfo {author} {\bibfnamefont {E.}~\bibnamefont
  {Lieb}}\ and\ \bibinfo {author} {\bibfnamefont {D.}~\bibnamefont {Mattis}},\
  }\href@noop {} {\bibfield  {journal} {\bibinfo  {journal} {J. Math. Phys.}\
  }\textbf {\bibinfo {volume} {3}},\ \bibinfo {pages} {749} (\bibinfo {year}
  {1962})}\BibitemShut {NoStop}%
\bibitem [{\citenamefont {Yang}\ and\ \citenamefont {Yang}(1966)}]{YANG66}%
  \BibitemOpen
  \bibfield  {author} {\bibinfo {author} {\bibfnamefont {C.~N.}\ \bibnamefont
  {Yang}}\ and\ \bibinfo {author} {\bibfnamefont {C.~P.}\ \bibnamefont
  {Yang}},\ }\href@noop {} {\bibfield  {journal} {\bibinfo  {journal} {Phys.
  Rev.}\ }\textbf {\bibinfo {volume} {150}},\ \bibinfo {pages} {321} (\bibinfo
  {year} {1966})}\BibitemShut {NoStop}%
\bibitem [{\citenamefont {Pratt}(1961)}]{PRAT61}%
  \BibitemOpen
  \bibfield  {author} {\bibinfo {author} {\bibfnamefont {G.~W.}\ \bibnamefont
  {Pratt}},\ }\href@noop {} {\bibfield  {journal} {\bibinfo  {journal} {Phys.
  Rev.}\ }\textbf {\bibinfo {volume} {122}},\ \bibinfo {pages} {489} (\bibinfo
  {year} {1961})}\BibitemShut {NoStop}%
\bibitem [{\citenamefont {Huang}(2000)}]{HUAN00}%
  \BibitemOpen
  \bibfield  {author} {\bibinfo {author} {\bibfnamefont {K.}~\bibnamefont
  {Huang}},\ }\href@noop {} {\emph {\bibinfo {title} {Statistical Mechanics}}}\
  (\bibinfo  {publisher} {John Wiley and Sons},\ \bibinfo {year}
  {2000})\BibitemShut {NoStop}%
\bibitem [{\citenamefont {Majlis}(2007)}]{MAJL07}%
  \BibitemOpen
  \bibfield  {author} {\bibinfo {author} {\bibfnamefont {N.}~\bibnamefont
  {Majlis}},\ }\href@noop {} {\emph {\bibinfo {title} {The Quantum Theory of
  Magnetism}}}\ (\bibinfo  {publisher} {World Scientific},\ \bibinfo {year}
  {2007})\BibitemShut {NoStop}%
\bibitem [{\citenamefont {Kuzemsky}(2010)}]{KUZE10}%
  \BibitemOpen
  \bibfield  {author} {\bibinfo {author} {\bibfnamefont {A.~L.}\ \bibnamefont
  {Kuzemsky}},\ }\href@noop {} {\bibfield  {journal} {\bibinfo  {journal} {Int.
  J. Mod. Phys. B}\ }\textbf {\bibinfo {volume} {24}},\ \bibinfo {pages} {835}
  (\bibinfo {year} {2010})}\BibitemShut {NoStop}%
\bibitem [{\citenamefont {Nagler}\ \emph {et~al.}(1991)\citenamefont {Nagler},
  \citenamefont {Tennant}, \citenamefont {Cowley}, \citenamefont {Perring},\
  and\ \citenamefont {Satija}}]{NAGL91}%
  \BibitemOpen
  \bibfield  {author} {\bibinfo {author} {\bibfnamefont {S.~E.}\ \bibnamefont
  {Nagler}}, \bibinfo {author} {\bibfnamefont {D.~A.}\ \bibnamefont {Tennant}},
  \bibinfo {author} {\bibfnamefont {R.~A.}\ \bibnamefont {Cowley}}, \bibinfo
  {author} {\bibfnamefont {T.~G.}\ \bibnamefont {Perring}}, \ and\ \bibinfo
  {author} {\bibfnamefont {S.~K.}\ \bibnamefont {Satija}},\ }\href@noop {}
  {\bibfield  {journal} {\bibinfo  {journal} {Phys. Rev. B}\ }\textbf {\bibinfo
  {volume} {44}},\ \bibinfo {pages} {12361} (\bibinfo {year}
  {1991})}\BibitemShut {NoStop}%
\bibitem [{\citenamefont {Motoyama}\ \emph {et~al.}(1996)\citenamefont
  {Motoyama}, \citenamefont {Eisaki},\ and\ \citenamefont {Uchida}}]{MOTO96}%
  \BibitemOpen
  \bibfield  {author} {\bibinfo {author} {\bibfnamefont {N.}~\bibnamefont
  {Motoyama}}, \bibinfo {author} {\bibfnamefont {H.}~\bibnamefont {Eisaki}}, \
  and\ \bibinfo {author} {\bibfnamefont {S.}~\bibnamefont {Uchida}},\
  }\href@noop {} {\bibfield  {journal} {\bibinfo  {journal} {Phys. Rev. Lett.}\
  }\textbf {\bibinfo {volume} {76}},\ \bibinfo {pages} {3212} (\bibinfo {year}
  {1996})}\BibitemShut {NoStop}%
\bibitem [{\citenamefont {Hirjibehedin}\ \emph {et~al.}(2006)\citenamefont
  {Hirjibehedin}, \citenamefont {Lutz},\ and\ \citenamefont
  {Heinrich}}]{HIRJ06}%
  \BibitemOpen
  \bibfield  {author} {\bibinfo {author} {\bibfnamefont {C.~F.}\ \bibnamefont
  {Hirjibehedin}}, \bibinfo {author} {\bibfnamefont {C.~P.}\ \bibnamefont
  {Lutz}}, \ and\ \bibinfo {author} {\bibfnamefont {A.~J.}\ \bibnamefont
  {Heinrich}},\ }\href@noop {} {\bibfield  {journal} {\bibinfo  {journal}
  {Science}\ }\textbf {\bibinfo {volume} {312}},\ \bibinfo {pages} {1021}
  (\bibinfo {year} {2006})}\BibitemShut {NoStop}%
\bibitem [{\citenamefont {Loth}\ \emph {et~al.}(2012)\citenamefont {Loth},
  \citenamefont {Baumann}, \citenamefont {Lutz}, \citenamefont {Eigler},\ and\
  \citenamefont {Heinrich}}]{LOTH12}%
  \BibitemOpen
  \bibfield  {author} {\bibinfo {author} {\bibfnamefont {S.}~\bibnamefont
  {Loth}}, \bibinfo {author} {\bibfnamefont {S.}~\bibnamefont {Baumann}},
  \bibinfo {author} {\bibfnamefont {C.~P.}\ \bibnamefont {Lutz}}, \bibinfo
  {author} {\bibfnamefont {D.~M.}\ \bibnamefont {Eigler}}, \ and\ \bibinfo
  {author} {\bibfnamefont {A.~J.}\ \bibnamefont {Heinrich}},\ }\href@noop {}
  {\bibfield  {journal} {\bibinfo  {journal} {Science}\ }\textbf {\bibinfo
  {volume} {335}},\ \bibinfo {pages} {196} (\bibinfo {year}
  {2012})}\BibitemShut {NoStop}%
\bibitem [{\citenamefont {Khajetoorians}\ \emph {et~al.}(2012)\citenamefont
  {Khajetoorians}, \citenamefont {Wiebe}, \citenamefont {Chilian},
  \citenamefont {Lounis}, \citenamefont {Bl{\"u}gel},\ and\ \citenamefont
  {Wiesendanger}}]{KHAJ12}%
  \BibitemOpen
  \bibfield  {author} {\bibinfo {author} {\bibfnamefont {A.~A.}\ \bibnamefont
  {Khajetoorians}}, \bibinfo {author} {\bibfnamefont {J.}~\bibnamefont
  {Wiebe}}, \bibinfo {author} {\bibfnamefont {B.}~\bibnamefont {Chilian}},
  \bibinfo {author} {\bibfnamefont {S.}~\bibnamefont {Lounis}}, \bibinfo
  {author} {\bibfnamefont {S.}~\bibnamefont {Bl{\"u}gel}}, \ and\ \bibinfo
  {author} {\bibfnamefont {R.}~\bibnamefont {Wiesendanger}},\ }\href@noop {}
  {\bibfield  {journal} {\bibinfo  {journal} {Nat. Phys.}\ }\textbf {\bibinfo
  {volume} {8}},\ \bibinfo {pages} {497} (\bibinfo {year} {2012})}\BibitemShut
  {NoStop}%
\bibitem [{\citenamefont {Holzberger}\ \emph {et~al.}(2013)\citenamefont
  {Holzberger}, \citenamefont {Schuh}, \citenamefont {Bl{\"u}gel},
  \citenamefont {Lounis},\ and\ \citenamefont {Wulfhekel}}]{HOLZ13}%
  \BibitemOpen
  \bibfield  {author} {\bibinfo {author} {\bibfnamefont {S.}~\bibnamefont
  {Holzberger}}, \bibinfo {author} {\bibfnamefont {T.}~\bibnamefont {Schuh}},
  \bibinfo {author} {\bibfnamefont {S.}~\bibnamefont {Bl{\"u}gel}}, \bibinfo
  {author} {\bibfnamefont {S.}~\bibnamefont {Lounis}}, \ and\ \bibinfo {author}
  {\bibfnamefont {W.}~\bibnamefont {Wulfhekel}},\ }\href@noop {} {\bibfield
  {journal} {\bibinfo  {journal} {Phys. Rev. Lett.}\ }\textbf {\bibinfo
  {volume} {110}},\ \bibinfo {pages} {157206} (\bibinfo {year}
  {2013})}\BibitemShut {NoStop}%
\bibitem [{\citenamefont {Spinelli}\ \emph {et~al.}(2014)\citenamefont
  {Spinelli}, \citenamefont {Bryant}, \citenamefont {Delgado}, \citenamefont
  {Fern{\'a}ndez-Rossier},\ and\ \citenamefont {Otte}}]{SPIN14}%
  \BibitemOpen
  \bibfield  {author} {\bibinfo {author} {\bibfnamefont {A.}~\bibnamefont
  {Spinelli}}, \bibinfo {author} {\bibfnamefont {B.}~\bibnamefont {Bryant}},
  \bibinfo {author} {\bibfnamefont {F.}~\bibnamefont {Delgado}}, \bibinfo
  {author} {\bibfnamefont {J.}~\bibnamefont {Fern{\'a}ndez-Rossier}}, \ and\
  \bibinfo {author} {\bibfnamefont {A.~F.}\ \bibnamefont {Otte}},\ }\href@noop
  {} {\bibfield  {journal} {\bibinfo  {journal} {Nat. Mater.}\ }\textbf
  {\bibinfo {volume} {13}},\ \bibinfo {pages} {782} (\bibinfo {year}
  {2014})}\BibitemShut {NoStop}%
\bibitem [{\citenamefont {Friedenauer}\ \emph {et~al.}(2008)\citenamefont
  {Friedenauer}, \citenamefont {Schmitz}, \citenamefont {Glueckert},
  \citenamefont {Porras},\ and\ \citenamefont {Sch{\"a}tz}}]{FRIE08}%
  \BibitemOpen
  \bibfield  {author} {\bibinfo {author} {\bibfnamefont {A.}~\bibnamefont
  {Friedenauer}}, \bibinfo {author} {\bibfnamefont {H.}~\bibnamefont
  {Schmitz}}, \bibinfo {author} {\bibfnamefont {J.~T.}\ \bibnamefont
  {Glueckert}}, \bibinfo {author} {\bibfnamefont {D.}~\bibnamefont {Porras}}, \
  and\ \bibinfo {author} {\bibfnamefont {T.}~\bibnamefont {Sch{\"a}tz}},\
  }\href@noop {} {\bibfield  {journal} {\bibinfo  {journal} {Nat. Phys.}\
  }\textbf {\bibinfo {volume} {4}},\ \bibinfo {pages} {757} (\bibinfo {year}
  {2008})}\BibitemShut {NoStop}%
\bibitem [{\citenamefont {Richerme}\ \emph {et~al.}(2014)\citenamefont
  {Richerme}, \citenamefont {Gong}, \citenamefont {Lee}, \citenamefont {Senko},
  \citenamefont {Smith}, \citenamefont {Foss-Feig}, \citenamefont {Michalakis},
  \citenamefont {Gorshkov},\ and\ \citenamefont {Monroe}}]{RICH14}%
  \BibitemOpen
  \bibfield  {author} {\bibinfo {author} {\bibfnamefont {P.}~\bibnamefont
  {Richerme}}, \bibinfo {author} {\bibfnamefont {Z.-X.}\ \bibnamefont {Gong}},
  \bibinfo {author} {\bibfnamefont {A.}~\bibnamefont {Lee}}, \bibinfo {author}
  {\bibfnamefont {C.}~\bibnamefont {Senko}}, \bibinfo {author} {\bibfnamefont
  {J.}~\bibnamefont {Smith}}, \bibinfo {author} {\bibfnamefont
  {M.}~\bibnamefont {Foss-Feig}}, \bibinfo {author} {\bibfnamefont
  {S.}~\bibnamefont {Michalakis}}, \bibinfo {author} {\bibfnamefont {A.~V.}\
  \bibnamefont {Gorshkov}}, \ and\ \bibinfo {author} {\bibfnamefont
  {C.}~\bibnamefont {Monroe}},\ }\href@noop {} {\bibfield  {journal} {\bibinfo
  {journal} {Nature}\ }\textbf {\bibinfo {volume} {511}},\ \bibinfo {pages}
  {198} (\bibinfo {year} {2014})}\BibitemShut {NoStop}%
\bibitem [{\citenamefont {Trotzky}\ \emph {et~al.}(2008)\citenamefont
  {Trotzky}, \citenamefont {Cheinet}, \citenamefont {F{\"o}lling},
  \citenamefont {Feld}, \citenamefont {Schnorrberger}, \citenamefont {Rey},
  \citenamefont {Polkovnikov}, \citenamefont {Demler}, \citenamefont {Lukin},\
  and\ \citenamefont {Bloch}}]{TROT08}%
  \BibitemOpen
  \bibfield  {author} {\bibinfo {author} {\bibfnamefont {S.}~\bibnamefont
  {Trotzky}}, \bibinfo {author} {\bibfnamefont {P.}~\bibnamefont {Cheinet}},
  \bibinfo {author} {\bibfnamefont {S.}~\bibnamefont {F{\"o}lling}}, \bibinfo
  {author} {\bibfnamefont {M.}~\bibnamefont {Feld}}, \bibinfo {author}
  {\bibfnamefont {U.}~\bibnamefont {Schnorrberger}}, \bibinfo {author}
  {\bibfnamefont {A.~M.}\ \bibnamefont {Rey}}, \bibinfo {author} {\bibfnamefont
  {A.}~\bibnamefont {Polkovnikov}}, \bibinfo {author} {\bibfnamefont {E.~A.}\
  \bibnamefont {Demler}}, \bibinfo {author} {\bibfnamefont {M.~D.}\
  \bibnamefont {Lukin}}, \ and\ \bibinfo {author} {\bibfnamefont
  {I.}~\bibnamefont {Bloch}},\ }\href@noop {} {\bibfield  {journal} {\bibinfo
  {journal} {Science}\ }\textbf {\bibinfo {volume} {319}},\ \bibinfo {pages}
  {295} (\bibinfo {year} {2008})}\BibitemShut {NoStop}%
\bibitem [{\citenamefont {Simon}\ \emph {et~al.}(2011)\citenamefont {Simon},
  \citenamefont {Bakr}, \citenamefont {Ma}, \citenamefont {Tai}, \citenamefont
  {Preiss},\ and\ \citenamefont {Greiner}}]{SIMO11}%
  \BibitemOpen
  \bibfield  {author} {\bibinfo {author} {\bibfnamefont {J.}~\bibnamefont
  {Simon}}, \bibinfo {author} {\bibfnamefont {W.~S.}\ \bibnamefont {Bakr}},
  \bibinfo {author} {\bibfnamefont {R.}~\bibnamefont {Ma}}, \bibinfo {author}
  {\bibfnamefont {M.~E.}\ \bibnamefont {Tai}}, \bibinfo {author} {\bibfnamefont
  {P.~M.}\ \bibnamefont {Preiss}}, \ and\ \bibinfo {author} {\bibfnamefont
  {M.}~\bibnamefont {Greiner}},\ }\href@noop {} {\bibfield  {journal} {\bibinfo
   {journal} {Nature}\ }\textbf {\bibinfo {volume} {472}},\ \bibinfo {pages}
  {307} (\bibinfo {year} {2011})}\BibitemShut {NoStop}%
\bibitem [{\citenamefont {Katsnelson}\ \emph {et~al.}(2001)\citenamefont
  {Katsnelson}, \citenamefont {Dobrovitski},\ and\ \citenamefont
  {Harmon}}]{KATS01}%
  \BibitemOpen
  \bibfield  {author} {\bibinfo {author} {\bibfnamefont {M.~I.}\ \bibnamefont
  {Katsnelson}}, \bibinfo {author} {\bibfnamefont {V.~V.}\ \bibnamefont
  {Dobrovitski}}, \ and\ \bibinfo {author} {\bibfnamefont {B.~N.}\ \bibnamefont
  {Harmon}},\ }\href@noop {} {\bibfield  {journal} {\bibinfo  {journal} {Phys.
  Rev. B}\ }\textbf {\bibinfo {volume} {63}},\ \bibinfo {pages} {212404}
  (\bibinfo {year} {2001})}\BibitemShut {NoStop}%
\bibitem [{\citenamefont {{Von Neumann}}(1955)}]{NEUM55}%
  \BibitemOpen
  \bibfield  {author} {\bibinfo {author} {\bibfnamefont {J.}~\bibnamefont {{Von
  Neumann}}},\ }\href@noop {} {\emph {\bibinfo {title} {Mathematical
  Foundations of Quantum Mechanics}}}\ (\bibinfo  {publisher} {Princeton
  University Press},\ \bibinfo {year} {1955})\BibitemShut {NoStop}%
\bibitem [{\citenamefont {Zurek}(2003)}]{ZURE03}%
  \BibitemOpen
  \bibfield  {author} {\bibinfo {author} {\bibfnamefont {W.~H.}\ \bibnamefont
  {Zurek}},\ }\href@noop {} {\bibfield  {journal} {\bibinfo  {journal} {Rev.
  Mod. Phys.}\ }\textbf {\bibinfo {volume} {75}},\ \bibinfo {pages} {715}
  (\bibinfo {year} {2003})}\BibitemShut {NoStop}%
\bibitem [{\citenamefont {Joos}\ \emph {et~al.}(2013)\citenamefont {Joos},
  \citenamefont {Zeh}, \citenamefont {Kiefer}, \citenamefont {Giulini},
  \citenamefont {Kupsch},\ and\ \citenamefont {Stamatescu}}]{JOOS13}%
  \BibitemOpen
  \bibfield  {author} {\bibinfo {author} {\bibfnamefont {E.}~\bibnamefont
  {Joos}}, \bibinfo {author} {\bibfnamefont {H.~D.}\ \bibnamefont {Zeh}},
  \bibinfo {author} {\bibfnamefont {C.}~\bibnamefont {Kiefer}}, \bibinfo
  {author} {\bibfnamefont {D.~J.~W.}\ \bibnamefont {Giulini}}, \bibinfo
  {author} {\bibfnamefont {J.}~\bibnamefont {Kupsch}}, \ and\ \bibinfo {author}
  {\bibfnamefont {I.~O.}\ \bibnamefont {Stamatescu}},\ }\href@noop {} {\emph
  {\bibinfo {title} {Decoherence and the Appearance of a Classical World in
  Quantum Theory}}}\ (\bibinfo  {publisher} {Springer Berlin Heidelberg},\
  \bibinfo {year} {2013})\BibitemShut {NoStop}%
\bibitem [{\citenamefont {Allahverdyan}\ \emph {et~al.}(2013)\citenamefont
  {Allahverdyan}, \citenamefont {Balian},\ and\ \citenamefont
  {Nieuwenhuizen}}]{NIEU13}%
  \BibitemOpen
  \bibfield  {author} {\bibinfo {author} {\bibfnamefont {A.~E.}\ \bibnamefont
  {Allahverdyan}}, \bibinfo {author} {\bibfnamefont {R.}~\bibnamefont
  {Balian}}, \ and\ \bibinfo {author} {\bibfnamefont {T.~M.}\ \bibnamefont
  {Nieuwenhuizen}},\ }\href@noop {} {\bibfield  {journal} {\bibinfo  {journal}
  {Phys. Rep.}\ }\textbf {\bibinfo {volume} {525}},\ \bibinfo {pages} {1 }
  (\bibinfo {year} {2013})}\BibitemShut {NoStop}%
\bibitem [{\citenamefont {Katsnelson}\ \emph {et~al.}(2000)\citenamefont
  {Katsnelson}, \citenamefont {Dobrovitski},\ and\ \citenamefont
  {Harmon}}]{KATS00}%
  \BibitemOpen
  \bibfield  {author} {\bibinfo {author} {\bibfnamefont {M.~I.}\ \bibnamefont
  {Katsnelson}}, \bibinfo {author} {\bibfnamefont {V.~V.}\ \bibnamefont
  {Dobrovitski}}, \ and\ \bibinfo {author} {\bibfnamefont {B.~N.}\ \bibnamefont
  {Harmon}},\ }\href@noop {} {\bibfield  {journal} {\bibinfo  {journal} {Phys.
  Rev. A}\ }\textbf {\bibinfo {volume} {62}},\ \bibinfo {pages} {022118}
  (\bibinfo {year} {2000})}\BibitemShut {NoStop}%
\bibitem [{\citenamefont {Hamieh}\ and\ \citenamefont
  {Katsnelson}(2005)}]{HAMI05}%
  \BibitemOpen
  \bibfield  {author} {\bibinfo {author} {\bibfnamefont {S.~D.}\ \bibnamefont
  {Hamieh}}\ and\ \bibinfo {author} {\bibfnamefont {M.~I.}\ \bibnamefont
  {Katsnelson}},\ }\href@noop {} {\bibfield  {journal} {\bibinfo  {journal}
  {Phys. Rev. A}\ }\textbf {\bibinfo {volume} {72}},\ \bibinfo {pages} {032316}
  (\bibinfo {year} {2005})}\BibitemShut {NoStop}%
\bibitem [{\citenamefont {Fano}(1957)}]{FANO57}%
  \BibitemOpen
  \bibfield  {author} {\bibinfo {author} {\bibfnamefont {U.}~\bibnamefont
  {Fano}},\ }\href@noop {} {\bibfield  {journal} {\bibinfo  {journal} {Rev.
  Mod. Phys.}\ }\textbf {\bibinfo {volume} {29}},\ \bibinfo {pages} {74}
  (\bibinfo {year} {1957})}\BibitemShut {NoStop}%
\bibitem [{\citenamefont {{Golub}}\ and\ \citenamefont {{Van
  Loan}}(1996)}]{GOLU96}%
  \BibitemOpen
  \bibfield  {author} {\bibinfo {author} {\bibfnamefont {G.~H.}\ \bibnamefont
  {{Golub}}}\ and\ \bibinfo {author} {\bibfnamefont {C.~F.}\ \bibnamefont {{Van
  Loan}}},\ }\href@noop {} {\emph {\bibinfo {title} {{Matrix Computations}}}}\
  (\bibinfo  {publisher} {John Hopkins University Press},\ \bibinfo {address}
  {Baltimore},\ \bibinfo {year} {1996})\BibitemShut {NoStop}%
\bibitem [{\citenamefont {Orbach}(1958)}]{ORBA58}%
  \BibitemOpen
  \bibfield  {author} {\bibinfo {author} {\bibfnamefont {R.}~\bibnamefont
  {Orbach}},\ }\href@noop {} {\bibfield  {journal} {\bibinfo  {journal} {Phys.
  Rev.}\ }\textbf {\bibinfo {volume} {112}},\ \bibinfo {pages} {309} (\bibinfo
  {year} {1958})}\BibitemShut {NoStop}%
\bibitem [{\citenamefont {Dobrovitski}\ and\ \citenamefont {{De
  Raedt}}(2003)}]{DOBR03}%
  \BibitemOpen
  \bibfield  {author} {\bibinfo {author} {\bibfnamefont {V.~V.}\ \bibnamefont
  {Dobrovitski}}\ and\ \bibinfo {author} {\bibfnamefont {H.~A.}\ \bibnamefont
  {{De Raedt}}},\ }\href@noop {} {\bibfield  {journal} {\bibinfo  {journal}
  {Phys. Rev. E}\ }\textbf {\bibinfo {volume} {67}},\ \bibinfo {pages} {056702}
  (\bibinfo {year} {2003})}\BibitemShut {NoStop}%
\bibitem [{\citenamefont {{De Raedt}}\ and\ \citenamefont
  {Michielsen}(2006)}]{RAED06}%
  \BibitemOpen
  \bibfield  {author} {\bibinfo {author} {\bibfnamefont {H.}~\bibnamefont {{De
  Raedt}}}\ and\ \bibinfo {author} {\bibfnamefont {K.}~\bibnamefont
  {Michielsen}},\ }\enquote {\bibinfo {title} {Computational methods for
  simulating quantum computers},}\ in\ \href@noop {} {\emph {\bibinfo
  {booktitle} {Quantum and molecular computing, quantum simulations}}},\
  \bibinfo {series} {Handbook of Theoretical and Computational Nanotechnology},
  Vol.~\bibinfo {volume} {3},\ \bibinfo {editor} {edited by\ \bibinfo {editor}
  {\bibfnamefont {M.}~\bibnamefont {Rieth}}\ and\ \bibinfo {editor}
  {\bibfnamefont {W.}~\bibnamefont {Schommers}}}\ (\bibinfo  {publisher}
  {American Scientific Publishers},\ \bibinfo {year} {2006})\ Chap.~\bibinfo
  {chapter} {1}, pp.\ \bibinfo {pages} {2--48}\BibitemShut {NoStop}%
\bibitem [{\citenamefont {Murmann}\ \emph {et~al.}(2015)\citenamefont
  {Murmann}, \citenamefont {Deuretzbacher}, \citenamefont {Z{\"u}rn},
  \citenamefont {Bjerlin}, \citenamefont {Reimann}, \citenamefont {Santos},
  \citenamefont {Lompe},\ and\ \citenamefont {Jochim}}]{MURM15}%
  \BibitemOpen
  \bibfield  {author} {\bibinfo {author} {\bibfnamefont {S.}~\bibnamefont
  {Murmann}}, \bibinfo {author} {\bibfnamefont {F.}~\bibnamefont
  {Deuretzbacher}}, \bibinfo {author} {\bibfnamefont {G.}~\bibnamefont
  {Z{\"u}rn}}, \bibinfo {author} {\bibfnamefont {J.}~\bibnamefont {Bjerlin}},
  \bibinfo {author} {\bibfnamefont {S.~M.}\ \bibnamefont {Reimann}}, \bibinfo
  {author} {\bibfnamefont {L.}~\bibnamefont {Santos}}, \bibinfo {author}
  {\bibfnamefont {T.}~\bibnamefont {Lompe}}, \ and\ \bibinfo {author}
  {\bibfnamefont {S.}~\bibnamefont {Jochim}},\ }\href@noop {} {\bibfield
  {journal} {\bibinfo  {journal} {Phys. Rev. Lett.}\ }\textbf {\bibinfo
  {volume} {115}},\ \bibinfo {pages} {215301} (\bibinfo {year}
  {2015})}\BibitemShut {NoStop}%
\bibitem [{\citenamefont {Bogoliubov}\ \emph {et~al.}(1986)\citenamefont
  {Bogoliubov}, \citenamefont {Izergin},\ and\ \citenamefont
  {Korepin}}]{BOGO86}%
  \BibitemOpen
  \bibfield  {author} {\bibinfo {author} {\bibfnamefont {N.~M.}\ \bibnamefont
  {Bogoliubov}}, \bibinfo {author} {\bibfnamefont {A.~G.}\ \bibnamefont
  {Izergin}}, \ and\ \bibinfo {author} {\bibfnamefont {V.~E.}\ \bibnamefont
  {Korepin}},\ }\href@noop {} {\bibfield  {journal} {\bibinfo  {journal} {Nucl.
  Phys. B}\ }\textbf {\bibinfo {volume} {275}},\ \bibinfo {pages} {687}
  (\bibinfo {year} {1986})}\BibitemShut {NoStop}%
\bibitem [{\citenamefont {Parkinson}\ and\ \citenamefont
  {Farnell}(2010)}]{PARK10}%
  \BibitemOpen
  \bibfield  {author} {\bibinfo {author} {\bibfnamefont {J.~B.}\ \bibnamefont
  {Parkinson}}\ and\ \bibinfo {author} {\bibfnamefont {D.~J.~J.}\ \bibnamefont
  {Farnell}},\ }\href@noop {} {\emph {\bibinfo {title} {{An Introduction to
  Quantum Spin Systems}}}},\ Vol.\ \bibinfo {volume} {816}\ (\bibinfo
  {publisher} {Springer},\ \bibinfo {year} {2010})\BibitemShut {NoStop}%
\bibitem [{\citenamefont {Landau}\ and\ \citenamefont
  {Lifshitz}(1984)}]{LAND84}%
  \BibitemOpen
  \bibfield  {author} {\bibinfo {author} {\bibfnamefont {L.~D.}\ \bibnamefont
  {Landau}}\ and\ \bibinfo {author} {\bibfnamefont {E.~M.}\ \bibnamefont
  {Lifshitz}},\ }\href@noop {} {\emph {\bibinfo {title} {Electrodynamics of
  Continuous Media}}},\ \bibinfo {edition} {2nd}\ ed.,\ Vol.~\bibinfo {volume}
  {8}\ (\bibinfo  {publisher} {Elsevier Science},\ \bibinfo {year}
  {1984})\BibitemShut {NoStop}%
\bibitem [{\citenamefont {Anderson}(1952)}]{ANDE52}%
  \BibitemOpen
  \bibfield  {author} {\bibinfo {author} {\bibfnamefont {P.~W.}\ \bibnamefont
  {Anderson}},\ }\href@noop {} {\bibfield  {journal} {\bibinfo  {journal}
  {Phys. Rev.}\ }\textbf {\bibinfo {volume} {86}},\ \bibinfo {pages} {694}
  (\bibinfo {year} {1952})}\BibitemShut {NoStop}%
\bibitem [{\citenamefont {{Hahn}}\ and\ \citenamefont {{Fine}}(2016)}]{HAHN16}%
  \BibitemOpen
  \bibfield  {author} {\bibinfo {author} {\bibfnamefont {W.}~\bibnamefont
  {{Hahn}}}\ and\ \bibinfo {author} {\bibfnamefont {B.~V.}\ \bibnamefont
  {{Fine}}},\ }\href@noop {} {\bibfield  {journal} {\bibinfo  {journal} {ArXiv
  e-prints}\ } (\bibinfo {year} {2016})},\ \Eprint
  {http://arxiv.org/abs/1601.06402} {arXiv:1601.06402 [quant-ph]} \BibitemShut
  {NoStop}%
\bibitem [{\citenamefont {Yi}\ and\ \citenamefont {Kim}(2013)}]{YI13}%
  \BibitemOpen
  \bibfield  {author} {\bibinfo {author} {\bibfnamefont {J.}~\bibnamefont
  {Yi}}\ and\ \bibinfo {author} {\bibfnamefont {Y.~W.}\ \bibnamefont {Kim}},\
  }\href@noop {} {\bibfield  {journal} {\bibinfo  {journal} {Phys. Rev. E}\
  }\textbf {\bibinfo {volume} {88}},\ \bibinfo {pages} {032105} (\bibinfo
  {year} {2013})}\BibitemShut {NoStop}%
\bibitem [{\citenamefont {Misra}\ and\ \citenamefont
  {Sudarshan}(1977)}]{MISR77}%
  \BibitemOpen
  \bibfield  {author} {\bibinfo {author} {\bibfnamefont {B.}~\bibnamefont
  {Misra}}\ and\ \bibinfo {author} {\bibfnamefont {E.~C.~G.}\ \bibnamefont
  {Sudarshan}},\ }\href@noop {} {\bibfield  {journal} {\bibinfo  {journal} {J.
  Math. Phys.}\ }\textbf {\bibinfo {volume} {18}},\ \bibinfo {pages} {756}
  (\bibinfo {year} {1977})}\BibitemShut {NoStop}%
\bibitem [{\citenamefont {Joos}(2009)}]{JOOS09}%
  \BibitemOpen
  \bibfield  {author} {\bibinfo {author} {\bibfnamefont {E.}~\bibnamefont
  {Joos}},\ }in\ \href@noop {} {\emph {\bibinfo {booktitle} {Compendium of
  Quantum Physics}}},\ \bibinfo {editor} {edited by\ \bibinfo {editor}
  {\bibfnamefont {D.}~\bibnamefont {Greenberger}}, \bibinfo {editor}
  {\bibfnamefont {K.}~\bibnamefont {Hentschel}}, \ and\ \bibinfo {editor}
  {\bibfnamefont {F.}~\bibnamefont {Weinert}}}\ (\bibinfo  {publisher}
  {Springer Berlin Heidelberg},\ \bibinfo {year} {2009})\ pp.\ \bibinfo {pages}
  {622--625}\BibitemShut {NoStop}%
\bibitem [{\citenamefont {Griffiths}(2005)}]{GRIFF05}%
  \BibitemOpen
  \bibfield  {author} {\bibinfo {author} {\bibfnamefont {D.~J.}\ \bibnamefont
  {Griffiths}},\ }\href@noop {} {\emph {\bibinfo {title} {Introduction to
  Quantum Mechanics}}},\ Pearson international edition\ (\bibinfo  {publisher}
  {Pearson Prentice Hall},\ \bibinfo {year} {2005})\BibitemShut {NoStop}%
\bibitem [{\citenamefont {Schwinger}(2001)}]{SCHW01}%
  \BibitemOpen
  \bibfield  {author} {\bibinfo {author} {\bibfnamefont {J.}~\bibnamefont
  {Schwinger}},\ }\href@noop {} {\emph {\bibinfo {title} {{Quantum Mechanics:
  Symbolism of Atomic Measurements}}}}\ (\bibinfo  {publisher} {Springer},\
  \bibinfo {year} {2001})\BibitemShut {NoStop}%
\end{thebibliography}%

\end{document}